\documentclass[12pt]{article}
\usepackage{epsfig}
\usepackage{algorithm,algorithmic}
\usepackage{amsmath}
\usepackage{amssymb}
\usepackage{url}
\usepackage{bigints}

\def\endproof{\hfill$\square$
\medskip}
\begin{document}
\title{\vspace{-17mm}
Elastic Shape Registration of Surfaces in 3D Space with Gradient Descent and Dynamic Programming}
\author{\tt\small Javier Bernal$^1$, Jim Lawrence$^{1,2}$\\
$^1${\small \sl National Institute of Standards and Technology,} \\
{\small \sl Gaithersburg, MD 20899, USA} \\
$^2${\small \sl George Mason University,} \\
{\small \sl 4400 University Dr, Fairfax, VA 22030, USA} \\
{\tt\small $\{$javier.bernal,james.lawrence$\}$} \\
{\tt\small @nist.gov \ \ \  lawrence@gmu.edu}}
\date{\ }
\maketitle
\vspace{-13mm}
\begin{abstract}
Algorithms based on gradient descent for computing the elastic shape registration of two simple
surfaces in $3-$dimensional space and therefore the elastic shape distance between them have been
proposed by Kurtek, Jermyn, et~al., and more recently by Riseth. Their algorithms are designed to
minimize a distance function between the surfaces by rotating and reparametrizing one of the surfaces,
the minimization for reparametrizing based on a gradient descent approach that may terminate at a
local solution. On the other hand, Bernal and Lawrence have proposed a similar algorithm, the
minimization for reparametrizing based on dynamic programming thus producing a partial not
necessarily optimal elastic shape registration of the surfaces. Accordingly, Bernal and Lawrence
have proposed to use the rotation and reparametrization computed with their algorithm as the
initial solution to any algorithm based on a gradient descent approach for reparametrizing.
Here we present results from doing exactly that. We also describe and justify the gradient
descent approach that is used for reparametrizing one of the~surfaces.
\\[0.2cm]
\end{abstract}
\section{\large Introduction}
\begin{figure}
\centering
\begin{tabular}{cc}
\includegraphics[width=0.4\textwidth]{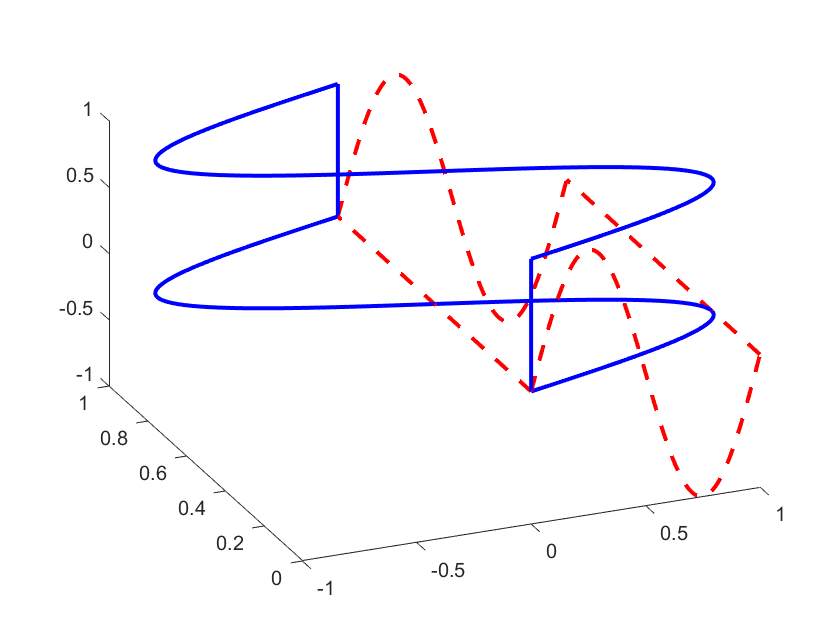}
&
\includegraphics[width=0.4\textwidth]{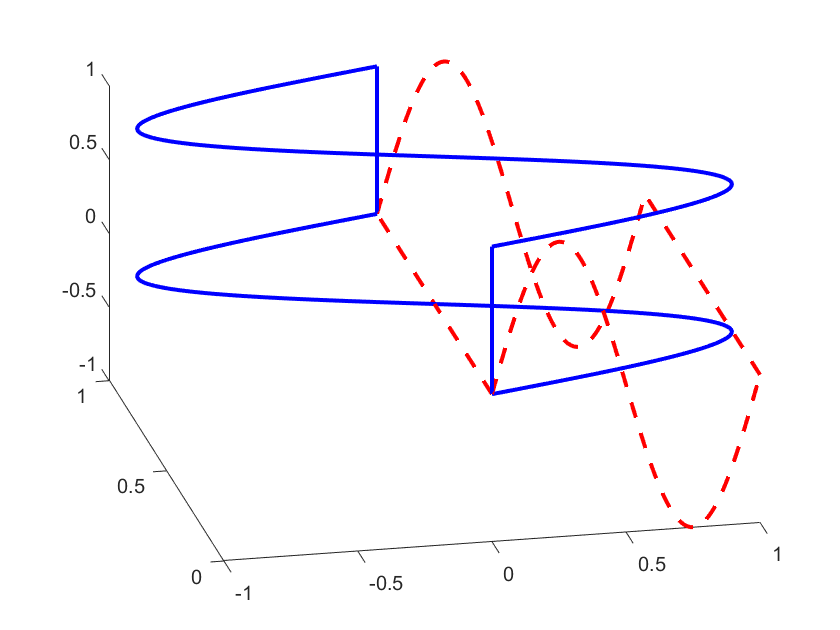}
\end{tabular}
\caption{\label{F:surfaces0}
Views of the boundaries of two surfaces in 3D space of identical sinusoidal shapes
so that the elastic shape distance between them is~zero.
}
\end{figure}

In this paper, we present results from computing the elastic shape registration of two simple surfaces
in $3-$dimensional space and the elastic shape distance between them with an algorithm based on a
gradient descent approach for reparametrizing one of the surfaces similar to those in
\cite{jermyn,kurtek}, and more recently in \cite{riseth}, using as the input initial solution to the
algorithm the rotation and reparametrization computed with the algorithm based on dynamic programming
presented in~\cite{bernal5} for reparametrizing one of the surfaces to obtain a partial elastic shape
registration of the surfaces.
We note, the gradient descent approach used to obtain our results is a generalization to surfaces in
$3-$dimensional space of the gradient descent approach for reparametrizing one of two curves in the plane
when computing the elastic shape distance between them as presented in \cite{srivastava}.
For the sake of completeness, we describe and justify the approach for curves as it is done there,
and then present and justify its generalization to surfaces in $3-$dimensional space. This generalization
together with its justification was developed independently of similar
work in~\cite{jermyn,kurtek,riseth}.
\\ \smallskip\\
Given that $S_1$ and $S_2$ are the two surfaces under consideration, we assume they are 
\emph{simple}, that is, we assume that for $D = [0,1]\times [0,1]$ in the $xy$ plane ($\mathbb{R}^2$),
i.e., the unit square in the plane, one-to-one functions $c_1$ and $c_2$ of class $C^1$ exist,
$c_1:D\rightarrow \mathbb{R}^3$, $c_2:D\rightarrow \mathbb{R}^3$, such that $S_1=c_1(D)$
and $S_2=c_2(D)$. We then say that $c_1$ and $c_2$ \emph{parametrize} or are \emph{parametrizations}
of $S_1$ and $S_2$, respectively, and that $S_1$ and $S_2$ are \emph{parametrized surfaces}
relative to $c_1$ and $c_2$, respectively. In addition, given a surface $S$ in $3-$dimensional space
and one-to-one functions $c$, $p$ of class $C^1$, $c:D\rightarrow \mathbb{R}^3$,
$p:D\rightarrow \mathbb{R}^3$, $c(D)=S$, $p(D)=S$, so that $c$ and $p$ are parametrizations of $S$,
we say $p$ is a \emph{reparametrization} of $c$ or that $p$
\emph{reparametrizes} $S$ (given as an image of $c$), if $p=c\circ h$ for a diffeomorphism $h$
from $D$ onto $D$.
\\ \smallskip\\
The computation of the elastic shape registration of two surfaces in 3D space together with
the elastic shape distance between them has applications in the study of geological terrains, surfaces
of anatomical objects and structures such as facial surfaces, etc. Figure~\ref{F:surfaces0} shows the
boundaries (solid blue and dashed red) of two surfaces of sinusoidal shape. Their shapes are identical
so that the elastic shape distance between them is zero. Note, the $x-$, $y-$ and $z-$ axes in the
figure are not to scale relative to one another.
\section{\large The Shape Function of a Parametrized Surface}
In this section we recall the definition of the shape function of a parametrized surface in
3-dimensional space as introduced in~\cite{bernal5}. A similar definition has been presented
in \cite{bernal2,bernal4,joshi,srivastava,srivastava2} in the
context of the shape function of a parametrized curve in $d-$dimensional space, $d$ any
positive integer. Accordingly, in \cite{bernal2,bernal4,joshi,srivastava,srivastava2}, given
$\beta:[0,1]\rightarrow \mathbb{R}^d$ of class $C^1$, a parametrization of a curve
in~$\mathbb{R}^d$, the shape function $q$ of $\beta$, i.e., the shape function $q$ of the
curve that $\beta$ parametrizes, $q:[0,1]\rightarrow \mathbb{R}^d$, is
defined by $q(t)=\dot{\beta}(t)/\sqrt{||\dot{\beta}(t)||}$, $t\in [0,1]$ ($d-$dimensional~0 if
$\dot{\beta}(t)$ equals $d-$dimensional~0).
It follows then that $q$ is square integrable as
\[ \int_0^1 ||q(t)||^2 dt=\int_0^1 ||\dot{\beta}(t)/\sqrt{||\dot{\beta}(t)||}\,||^2 dt=
\int_0^1 ||\dot{\beta}(t)|| dt \]
which is the length of the curve that $\beta$ parametrizes,
where $||\cdot||$ is the $d-$dimensional Euclidean norm.
Again with $q$ the shape function of $\beta$ and $\Gamma$ the set
of orientation-preserving diffeomorphisms of $[0,1]$ so that for $\gamma\in\Gamma$ then
$\dot{\gamma}\geq 0$ on $[0,1]$, it then follows that for $\gamma\in\Gamma$ the shape function of
the reparametrization $\beta\circ\gamma$ of $\beta$ is
$(q,\gamma)=(q\circ\gamma)\sqrt{\dot{\gamma}}$.  With $||q||_2 = (\int_0^1 ||q(t)||^2dt)^{1/2}$,
we also note that given $\beta_1$, $\beta_2:[0,1]\rightarrow \mathbb{R}^d$ of class $C^1$,
parametrizations of curves in $\mathbb{R}^d$ with shape functions $q_1$, $q_2$, respectively,
then $||(q_1,\gamma) - (q_2,\gamma)||_2=||q_1-q_2||_2$ for any $\gamma\in\Gamma$, and from this, etc.,
with $\Gamma_0=\{\gamma\in\Gamma,\ \dot{\gamma}>0$ on $[0,1]\}$, $SO(d)$ the set of $d\times d$
rotation matrices, it has been demonstrated \cite{bernal2,dogan,srivastava} that the number
inf$_{R\in SO(d),\gamma\in\Gamma_0}||Rq_1-(q_2,\gamma)||_2$ can then be used as a well-defined distance
between the two curves that $\beta_1$,~$\beta_2$ parametrize, $\beta_1$ and $\beta_2$ both
normalized to parametrize curves of length~1.
\\ \smallskip\\
As for the definition of the shape function of a parametrized surface in 3-dimensional space,
again with $D = [0,1]\times [0,1]$ in the $xy$ plane ($\mathbb{R}^2$),
given a one-to-one function $c$ of class $C^1$, $c:D\rightarrow \mathbb{R}^3$,
so that for $(u,v)$ in $D$, $c$ takes $(u,v)$ to $c(u,v)$ in~$\mathbb{R}^3$, $c$
a parametrization of a surface $S$ in $3-$dimensional space, the shape function $q$ of $c$,
$q:D\rightarrow \mathbb{R}^3$, i.e., the shape function $q$ of the surface $S$ that $c$ parametrizes
is defined by \[ q(u,v)=(\frac{\partial c}{\partial u}(u,v) \times \frac{\partial c}{\partial v}(u,v))/
\sqrt{|| \frac{\partial c}{\partial u}(u,v) \times \frac{\partial c}{\partial v}(u,v)||} \]
($3-$dimensional~0 if $\frac{\partial c}{\partial u}(u,v) \times \frac{\partial c}{\partial v}(u,v)$
equals $3-$dimensional~0), where $||\cdot||$ is the $3-$dimensional Euclidean norm. It follows then
that $q$ is square integrable as
\[ \int\int_D ||q(u,v)||^2 du\,dv =
\int\int_D || \frac{\partial c}{\partial u}(u,v) \times \frac{\partial c}{\partial v}(u,v) ||\,du\,dv
\]
which is the surface area of~$S$.
\\ \smallskip\\
With $c$, $q$, $D$, $S$ as above, in a manner similar to the one described above in the
context of the shape function of the parametrization of a curve in $d-$dimensional space,
the shape function of a reparametrization of~$c$ can be computed from the shape function $q$ of~$c$.
With $p$ as a reparametrization of $c$, i.e., $p$ a parametrization of $S$ and $p=c\circ h$ for a
diffeomorphism $h$ from $D$ onto $D$, $h(r,t)=(h_1(r,t),h_2(r,t))$, assuming
$\frac{\partial h}{\partial (r,t)}\geq 0$ on~$D$, $\frac{\partial h}{\partial (r,t)}$ the determinant
of the Jacobian of~$h$, i.e., $\frac{\partial h}{\partial (r,t)} =
\frac{\partial h_1}{\partial r}\frac{\partial h_2}{\partial t}
-\frac{\partial h_1}{\partial t}\frac{\partial h_2}{\partial r}$,
and defining a function on $D$ into $\mathbb{R}^3$, which we denote by $(q,h)$,
\[ (q,h)\equiv (q\circ h)\sqrt{\frac{\partial h}{\partial (r,t)}}, \]
then as established in~\cite{bernal5} the shape function on $D$ of the reparametrization $p=c\circ h$
of $c$ is then~$(q,h)$.
\\ \smallskip\\
We note as well, given $D$, $h$, $\frac{\partial h}{\partial (r,t)}$ as above,
$\frac{\partial h}{\partial (r,t)}\geq 0$ on~$D$;
$S_1$, $S_2$ surfaces, $c_1$, $c_2$ parametrizations of $S_1$, $S_2$, respectively,
$p_1$, $p_2$ parametrizations of $S_1$, $S_2$, respectively,
$p_1$, $p_2$ reparametrizations of $c_1$, $c_2$, respectively, $p_1=c_1\circ h$, $p_2=c_2\circ h$;
$q_1$, $q_2$, $\hat{q}_1$,~$\hat{q}_2$ the shape functions of $c_1$, $c_2$, $p_1$, $p_2$, respectively,
then as established in~\cite{bernal5}, with $h(r,t)= (h_1(r,t),h_2(r,t))=(u(r,t),v(r,t))$,
\begin{eqnarray*}
||\hat{q}_1-\hat{q}_2||_2 &\equiv& \big(\int\int_D ||\hat{q}_1-\hat{q}_2)||^2dr\,dt\big)^{1/2}\\
&=& \big(\int\int_D ||q_1-q_2||^2du\,dv\big)^{1/2}\\
&\equiv& ||q_1-q_2||_2.
\end{eqnarray*}
\ \smallskip\\
Based on results about shape functions of parametrized surfaces such as the results above, and
using arguments similar to arguments for justifying the definition of the distance between
curves in $d-$dimensional space found in \cite{bernal2,dogan,srivastava}, it has been
demonstrated as pointed out in~\cite{bernal5} that with $D$ as above, given two simple surfaces
$S_1$ and $S_2$ parametrized by functions $c_1$ and $c_2$, $c_1:D\rightarrow \mathbb{R}^3$,
$c_2:D\rightarrow \mathbb{R}^3$, $S_1=c_1(D)$, $S_2=c_2(D)$, letting $SO(3)$ be the set of
$3\times 3$ rotation matrices, $\Sigma_0$ the set of all diffeomorphisms $h$ from $D$ onto $D$,
with $h(r,t)=(h_1(r,t),h_2(r,t))$, $\frac{\partial h}{\partial (r,t)}>0$ on~$D$,
$\frac{\partial h}{\partial (r,t)}$ the determinant of the Jacobian of~$h$,
and $q_1$ and $q_2$ the shape functions of $c_1$ and $c_2$, respectively,
the number $\mathrm{inf}_{R\in SO(3),h\in\Sigma_0}||Rq_1-(q_2,h)||_2$, i.e.,
\[ \mathrm{inf}_{R\in SO(3),h\in\Sigma_0}\big(\int\int_D ||Rq_1-(q_2,h)||^2dr\,dt\big)^{1/2} = \]
\[ \mathrm{inf}_{R\in SO(3),h\in\Sigma_0}\big(\int_0^1\int_0^1 \big|\big|Rq_1
-(q_2\circ h)\sqrt{\frac{\partial h}{\partial (r,t)}}\big|\big|^2dr\,dt\big)^{1/2} \]
can be used as a well-defined distance between the surfaces $S_1$ and~$S_2$,
$c_1$ and $c_2$ both normalized to parametrize surfaces of area equal to~1.
\section{\large Gradient Descent Optimization over the Group of Reparametrizations of a
Curve in the Plane}
In this section, for the sake of completeness, we describe and justify the gradient descent
approach in \cite{srivastava} in the same way it is done there, for reparametrizing one
of two curves in the plane when computing the elastic shape distance between them. In what
follows, $\Gamma_I$ will denote the set of functions~$\gamma:[0,1]\rightarrow [0,1]$, with
$\gamma(0)=0$, $\gamma(1)=1$, such that $\gamma^{-1}$, the inverse of $\gamma$, exists,
both $\gamma$ and $\gamma^{-1}$ are smooth, and $\dot{\gamma}>0$ on~$[0,1]$.
With $H:\Gamma_I\rightarrow\mathbb{R}_{\geq 0}$ defined by
\[H(\gamma)=\int_{0}^{1}||q_1(t)-q_2(\gamma(t))\sqrt{\dot{\gamma}(t)}||^2\;dt,\]
where $q_1$, $q_2$ are shape functions of parametrized curves in the plane of length~1, the goal
then is to find $\gamma\in\Gamma_I$ that minimizes $H(\gamma)$.
\\ \smallskip\\
In order to find any such $\gamma$ with a gradient approach, as illustrated below, a gradient of
$H$ is computed with respect to $\Gamma_I$ at the $k^{th}$ iteration of the approach, from which 
$\gamma_k$ and $\gamma^{(k)}$ in $\Gamma_I$ are then computed, so that inductively with $\gamma_0$
in $\Gamma_I$ as an initial solution, $\gamma_0$ possibly equal to the identity function
$\gamma_{id}$ in $\Gamma_I$, then $\gamma^{(k)}=\gamma_0\circ\gamma_1\circ \ldots\circ\gamma_k$,
and $H(\gamma^{(0)})>H(\gamma^{(1)})>\ldots>H(\gamma^{(k)})$. In reality, inductively, with
$\tilde{q}_2=(q_2\circ\gamma^{(k-1)})\sqrt{\dot{\gamma^{(k-1)}}}$, at the $k^{th}$ iteration of
the approach it is the gradient of 
\[H_k(\gamma)=\int_{0}^{1}||q_1(t)-\tilde{q}_2(\gamma(t))\sqrt{\dot{\gamma}(t)}||^2\;dt\]
that is actually computed, from which $\gamma_k$ is computed, and as verified below, with
$\gamma^{(k)}$ as above, then $H(\gamma^{(k)})=H_k(\gamma_k)$.
\\ \smallskip\\
With $[q_2]_{\Gamma_I}$ denoting the orbit of $q_2$, i.e.,
$[q_2]_{\Gamma_I}=\{\tilde{q}\;|\;\tilde{q}=(q_2\circ\gamma)\sqrt{\dot{\gamma}},\;\gamma\in\Gamma_I\}$,
and $\tilde{q}_2$ in $[q_2]_{\Gamma_I}$ given by
$\tilde{q}_2=(q_2\circ\gamma^{(k)})\sqrt{\dot{\gamma^{(k)}}}$, at the $(k+1)^{th}$ iteration
of the approach, letting $T_{\tilde{q}_2}([q_2]_{\Gamma_I})$ be the tangent space to
$[q_2]_{\Gamma_I}$ at $\tilde{q}_2$, $T_{\gamma_{id}}(\Gamma_I)$ the tangent space to $\Gamma_I$ at
$\gamma_{id}$, and $\phi$ the mapping from $\Gamma_I$ into $[q_2]_{\Gamma_I}$ defined by
$\phi(\gamma)=(\tilde{q}_2\circ\gamma)\sqrt{\dot{\gamma}}$, $\gamma\in\Gamma_I$,
since $\phi(\gamma_{id})=\tilde{q}_2$, then the differential
$d\phi_{\gamma_{id}}:T_{\gamma_{id}}(\Gamma_I)\rightarrow T_{\tilde{q}_2}([q_2]_{\Gamma_I})$
of $\phi$ at $\gamma_{id}$ can be defined. Accordingly, given $v$ in
$T_{\gamma_{id}}(\Gamma_I)$,
the lemma that follows shows how to compute $d\phi_{\gamma_{id}}(v)$, and using the result of this
computation, the theorem that also follows shows how to compute at the $(k+1)^{th}$ iteration of
the approach the directional derivative $\nabla_v H$ of $H$ (actually $\nabla_v H_{k+1}$ of
$H_{k+1}$) in the direction of $v$.
\\ \smallskip\\
We note then that given $v_i$, $i=1,2,3,\ldots$, an orthonormal basis of the vector space
$T_{\gamma_{id}}(\Gamma_I)$ under some metric, e.g., $\frac {1}{\sqrt{2}\pi n}\sin(2\pi nt)$,
$\frac {1}{\sqrt{2}\pi n}(\cos(2\pi nt)-1)$, $t\in [0,1]$, $n=1,2,3,\ldots$, under the Palais
metric, $\sqrt{2}\sin(m\pi t)$, $t\in [0,1]$, $m=1,2,3,\ldots$, under the $L^2$ metric,
at the $(k+1)^{th}$ iteration of the approach the gradient of $H_{k+1}$ is approximated
by $\nabla H_{k+1} = \sum_{i=1}^N (\nabla_{v_i} H_{k+1}) v_i$, for a large~$N$.
If under the same metric $\nabla H_{k+1}$ is considered to be small enough, then $\gamma^{(k)}$
is taken to be the solution of the gradient descent approach, although perhaps a local solution.
Otherwise, using a small step size $\delta >0$, $\gamma_{k+1}$ and $\gamma^{(k+1)}$
are computed, $\gamma_{k+1}=\gamma_{id}-\delta\;\nabla H_{k+1}$,
$\gamma^{(k+1)}=\gamma^{(k)}\circ\gamma_{k+1}$. That $\gamma^{(k+1)}$ is computed so that
$H(\gamma^{(k+1)})=H_{k+1}(\gamma_{k+1})$ follows by letting $\tilde{q}_2$ be as obtained
at the $k^{th}$ iteration of the approach, and proving that
\[ (q_2\circ\gamma^{(k+1)})\sqrt{\dot{\gamma^{(k+1)}}}
=(\tilde{q}_2\circ\gamma_{k+1})\sqrt{\dot{\gamma_{k+1}}}
\mathrm{\ \ as\ is\ done\ here\:\!\!\!}:\]
\begin{eqnarray*}
&\ &(\tilde{q}_2\circ\gamma_{k+1})\sqrt{\dot{\gamma_{k+1}}}
=((q_2\circ\gamma^{(k)})\sqrt{\dot{\gamma^{(k)}}})\circ\gamma_{k+1})\sqrt{\dot{\gamma_{k+1}}}\\
 &=&(q_2\circ\gamma^{(k)}\circ\gamma_{k+1})\sqrt{\dot{\gamma^{(k)}}\circ\gamma_{k+1}}\sqrt{\dot{\gamma_{k+1}}}\\
&=&(q_2\circ\gamma^{(k)}\circ\gamma_{k+1})\sqrt{(\dot{\gamma^{(k)}}\circ\gamma_{k+1})\dot{\gamma_{k+1}}}
=(q_2\circ\gamma^{(k+1)})\sqrt{\dot{\gamma^{(k+1)}}}.
\end{eqnarray*}
\ \\
The lemma and theorem follow. Here shape functions are $C^1$ as well.
\\ \smallskip\\
{\bf Lemma 1:} Given $q$, the shape function of a curve in the plane,
$\phi:\Gamma_I\rightarrow [q]_{\Gamma_I}$, $\phi(\gamma)=(q\circ\gamma)\sqrt{\dot{\gamma}}$, then given
$\gamma$ in $\Gamma_I$, $v$ in $T_{\gamma}(\Gamma_I)$, with $\dot{q}$ the Jacobian of $q$,
the differential of $\phi$ at $\gamma$ applied on $v$ is
\[ (d\phi_{\gamma}(v))(s)=\sqrt{\dot{\gamma}(s)}\dot{q}(\gamma(s))v(s)+
\frac{1}{2\sqrt{\dot{\gamma}(s)}}\dot{v}(s)q(\gamma(s)),\ s\in [0,1].\]
{\bf Proof:} Let $\alpha(\tau,\cdot)$ be a differentiable path in $\Gamma_I$ passing through $\gamma$ at
$\tau=0$, i.e., $\alpha(0,s) = \gamma(s)$, $s\in [0,1]$. Let the velocity of this path at $\tau=0$ be given
by $v\in T_{\gamma}(\Gamma_I)$, i.e., $v(s)= \frac{\partial\alpha}{\partial\tau}(0,s)$, $s\in [0,1]$. Note
as well that $\frac{\partial\alpha}{\partial s}(0,s) = \dot{\gamma}(s)$,
$\frac{\partial^2\alpha}{\partial s\partial\tau}(0,s) = \dot{v}(s)$, $s\in [0,1]$.
\\ \smallskip\\
Since $\alpha(\tau,\cdot)$ is a path in $\Gamma_I$, then $\phi(\alpha(\tau,\cdot))$ is the corresponding
path in $[q]_{\Gamma_I}$, and since $v$ is the velocity of $\alpha(\tau,\cdot)$ at $\tau=0$, then the
velocity of $\phi(\alpha(\tau,\cdot))$ at $\tau=0$ is $d\phi_{\gamma}(v)$, that is to say
\begin{eqnarray*}
	(d\phi_{\gamma}(v))(s)&=&\frac{\partial}{\partial\tau}|_{\tau=0}\;\phi(\alpha(\tau,s))
	=\frac{\partial}{\partial\tau}|_{\tau=0}\;\Big(\sqrt{\frac{\partial\alpha}{\partial s}(\tau,s)}
	\;q(\alpha(\tau,s))\Big)\\
	&=&\Big(\sqrt{\frac{\partial\alpha}{\partial s}(\tau,s)}
	\;\dot{q}(\alpha(\tau,s))\;\frac{\partial\alpha}{\partial\tau}(\tau,s)+\\
	&\ & \frac{1}{2\sqrt{\frac{\partial\alpha}{\partial s}(\tau,s)}}
	\;\frac{\partial^2\alpha}{\partial\tau\partial s}(\tau,s)\;q(\alpha(\tau,s))\Big)|_{\tau=0}\\
        &=&\sqrt{\dot{\gamma}(s)}\dot{q}(\gamma(s))v(s)+
	\frac{1}{2\sqrt{\dot{\gamma}(s)}}\dot{v}(s)q(\gamma(s)),\ s\in [0,1].
\end{eqnarray*}
\endproof\\
{\bf Corollary 1:} Given $q$, $\phi$, $v$ as above, then
\[ (d\phi_{\gamma_{id}}(v))(s)=\dot{q}(s)v(s)+ \frac{1}{2}\dot{v}(s)q(s),\ s\in [0,1].\]
{\bf Theorem 1:}
With $H_{k+1}(\gamma)=\int_{0}^{1}||q_1(t)-\tilde{q}_2(\gamma(t))\sqrt{\dot{\gamma}(t)}||^2 dt$
so that $\tilde{q}_2=(q_2\circ\gamma^{(k)})\sqrt{\dot{\gamma^{(k)}}}$ from the definition of $H_{k+1}$,
then the directional derivative of $H_{k+1}$ in the direction of $v\in T_{\gamma_{id}}(\Gamma_I)$ is
\[\nabla_v H_{k+1} = -2\int_0^1\Big\langle q_1(t)-\tilde{q}_2(t)\;,\;\dot{\tilde{q}}_2(t)v(t)+
\frac{1}{2}\tilde{q}_2(t)\dot{v}(t)\Big\rangle dt. \]
{\bf Proof:} Let $\alpha(\tau,\cdot)$ be a differentiable path in $\Gamma_I$ passing through
$\gamma_{id}$ at $\tau=0$, i.e., $\alpha(0,t)=\gamma_{id}(t)$, $t\in [0,1]$, with the velocity of this
path at $\tau=0$ equal to $v$, i.e., $\frac{\partial\alpha}{\partial\tau}(0,t)=v(t)$, $t\in [0,1]$.
\\ \smallskip\\
Note, $H_{k+1}(\alpha(\tau,t))$ equals
\[ \int_{0}^{1}\Big\langle q_1(t)-\tilde{q}_2(\alpha(\tau,t))\sqrt{\frac{\partial\alpha}{\partial t}
(\tau,t)}\;,\;
q_1(t)-\tilde{q}_2(\alpha(\tau,t))\sqrt{\frac{\partial\alpha}{\partial t}(\tau,t)}\Big\rangle dt.\]
Using the fact that in general
$\frac{d}{ds}\langle f(s),g(s)\rangle = \langle f(s),g'(s)\rangle+\langle f'(s),g(s)\rangle$, from
which $\frac{d}{ds}\langle f(s),f(s)\rangle = 2\langle f(s),f'(s)\rangle$,
then differentiating $H_{k+1}(\alpha(\tau,t))$ with respect to $\tau$ by differentiating this last
integral with respect to $\tau$ (done under the integral), and setting $\tau$ equal to zero, gives
$\nabla_v H_{k+1}$ equals
\[2\int_{0}^{1}\Big\langle q_1(t)-\tilde{q}_2(\gamma_{id}(t))\sqrt{\dot{\gamma_{id}}(t)}\;,\;
\frac{\partial}{\partial \tau}|_{\tau=0}\Big(-\tilde{q}_2(\alpha(\tau,t))
\sqrt{\frac{\partial\alpha}{\partial t}(\tau,t)}\Big)\Big\rangle dt\]
\[=-2\int_{0}^{1}\Big\langle q_1(t)-\tilde{q}_2(t)\;,\;
\frac{\partial}{\partial \tau}|_{\tau=0}\Big(\tilde{q}_2(\alpha(\tau,t))
\sqrt{\frac{\partial\alpha}{\partial t}(\tau,t)}\Big)\Big\rangle dt\]
\[=-2\int_0^1\Big\langle q_1(t)-\tilde{q}_2(t)\;,\;\dot{\tilde{q}}_2(t)v(t)+
\frac{1}{2}\tilde{q}_2(t)\dot{v}(t)\Big\rangle dt \]
by Corollary 1 and the proof of Lemma 1 as $\alpha$ here is the same as $\alpha$
there with $\gamma$ equal to~$\gamma_{id}$. \endproof
\\ \smallskip\\
Note that given $\gamma\in\Gamma_I$ and a differentiable path $\alpha(\tau,\cdot)$ in $\Gamma_I$ through
$\gamma$ at $\tau=0$, i.e., $\alpha(0,t) = \gamma(t)$, $t\in [0,1]$, then since for any real numbers
$\tau_1$, $\tau_2$, close or equal to 0,
$\alpha(\tau_1,0) - \alpha(\tau_2,0) = 0-0=0$ and
$\alpha(\tau_1,1) - \alpha(\tau_2,1) = 1-1=0$, 
it can also be shown that
\[T_{\gamma}(\Gamma_I)=\{v:[0,1]\rightarrow \mathbb{R}\;|\;v(0)=0,v(1)=0,\;v\ \mathrm{smooth}\}.\]
We note that under this identification of $T_{\gamma}(\Gamma_I)$, each
member of either of the sets of functions given above as examples of bases for
$T_{\gamma_{id}}(\Gamma_I)$ is indeed in~$T_{\gamma_{id}}(\Gamma_I)$.
\section{\large Gradient Descent Optimization over the Group of Reparametrizations of a
Surface in 3D Space}
In this section, inspired by ideas used in the previous section, we describe and justify
a gradient descent approach for reparametrizing one of two surfaces in $3-$dimensional space
when computing the elastic shape distance between them. The approach is a generalization to
surfaces in $3-$dimensional space of the gradient descent approach for reparametrizing one of two
curves in the plane when computing the elastic shape distance between them as presented in
\cite{srivastava} and in the previous section. This generalization together with its
justification was developed independently of similar work in~\cite{jermyn,kurtek,riseth}. Again
with $D$ the unit square in the plane, in what follows, $\Gamma_D$ will denote the set of functions
$h: D\rightarrow D$, with $h(r,t)=(h_1(r,t),h_2(r,t))$, $(r,t)\in D$, satisfying boundary conditions
$h_1(0,t)=0$, $h_2(r,0)=0$, $h_1(1,t)=1$, $h_2(r,1)=1$, such that $h^{-1}$, the inverse of $h$,
exists, both $h$ and $h^{-1}$ are smooth, and $\frac{\partial h}{\partial (r,t)}>0$ on~$D$,
$\frac{\partial h}{\partial (r,t)}$ the determinant of the Jacobian of~$h$, i.e.,
$\frac{\partial h}{\partial (r,t)} = \frac{\partial h_1}{\partial r}\frac{\partial h_2}{\partial t}
-\frac{\partial h_1}{\partial t}\frac{\partial h_2}{\partial r}$.
With $H:\Gamma_D\rightarrow\mathbb{R}_{\geq 0}$ defined by
\[ H(h) = \int_0^1\int_0^1 \Big|\Big|q_1(r,t)
-q_2(h(r,t))\sqrt{\frac{\partial h}{\partial (r,t)}(r,t)}\Big|\Big|^2dr\,dt, \]
where $q_1$, $q_2$ are shape functions of parametrized surfaces in $3-$dimensional space of surface
area~1, the goal then is to find $h\in \Gamma_D$ that minimizes~$H(h)$.
\\ \smallskip\\
Again, inspired by ideas used in the previous section, in order to find any such $h\in \Gamma_D$
with a gradient approach, as illustrated below, a gradient of
$H$ is computed with respect to $\Gamma_D$ at the $k^{th}$ iteration of the approach, from which 
$h_k$ and $h^{(k)}$ in $\Gamma_D$ are then computed, so that inductively with $h_0$ in $\Gamma_D$
as an initial solution, $h_0$ possibly equal to the identity function $h_{id}$ in $\Gamma_D$, or
possibly equal to another element of $\Gamma_D$ such as one computed with the algorithm based on
dynamic programming presented in~\cite{bernal5} that partially minimizes~$H(h)$,
then $h^{(k)}=h_0\circ h_1\circ \ldots\circ h_k$,
and $H(h^{(0)})>H(h^{(1)})>\ldots>H(h^{(k)})$. In reality, inductively, with
$\tilde{q}_2(r,t)=q_2(h^{(k-1)}(r,t))\sqrt{\frac{\partial h^{(k-1)}}{\partial (r,t)}(r,t)}$,
$(r,t)\in D$, at the $k^{th}$ iteration of the approach it is the gradient of 
\[ H_k(h) = \int_0^1\int_0^1 \Big|\Big|q_1(r,t)
-\tilde{q}_2(h(r,t))\sqrt{\frac{\partial h}{\partial (r,t)}(r,t)}\Big|\Big|^2dr\,dt \]
that is actually computed, from which $h_k$ is computed, and as verified below, with
$h^{(k)}$ as above, then $H(h^{(k)})=H_k(h_k)$.
\\ \smallskip\\
With $[q_2]_{\Gamma_D}$ denoting the orbit of $q_2$, i.e.,
\[[q_2]_{\Gamma_D}= \big\{\tilde{q}\;|\;\tilde{q}(r,t)=
q_2(h(r,t))\sqrt{\frac{\partial h}{\partial (r,t)}(r,t)},\; (r,t)\in D,\; h\in\Gamma_D\big\},\]
and $\tilde{q}_2$ in $[q_2]_{\Gamma_D}$ given by
$\tilde{q}_2(r,t)=q_2(h^{(k)}(r,t))\sqrt{\frac{\partial h^{(k)}}{\partial (r,t)}(r,t)}$,
at the $(k+1)^{th}$ iteration
of the approach, letting $T_{\tilde{q}_2}([q_2]_{\Gamma_D})$ be the tangent space to
$[q_2]_{\Gamma_D}$ at $\tilde{q}_2$, $T_{h_{id}}(\Gamma_D)$ the tangent space to $\Gamma_D$ at
$h_{id}$, and $\phi$ the mapping from $\Gamma_D$ into $[q_2]_{\Gamma_D}$ defined by
$\phi(h)(r,t)=\tilde{q}_2(h(r,t))\sqrt{\frac{\partial h}{\partial (r,t)}(r,t)}$, $(r,t)\in D$,
$h\in\Gamma_D$, since $\phi(h_{id})=\tilde{q}_2$, then the differential
$d\phi_{h_{id}}:T_{h_{id}}(\Gamma_D)\rightarrow T_{\tilde{q}_2}([q_2]_{\Gamma_D})$
of $\phi$ at $h_{id}$ can be defined. Accordingly, given $v$ in
$T_{h_{id}}(\Gamma_D)$,
the lemma that follows shows how to compute $d\phi_{h_{id}}(v)$, and using the result of this
computation, the theorem that also follows shows how to compute at the $(k+1)^{th}$ iteration of
the approach the directional derivative $\nabla_v H$ of $H$ (actually $\nabla_v H_{k+1}$ of
$H_{k+1}$) in the direction of $v$.
\\ \smallskip\\
We note then that given $w_i$, $i=1,2,3,\ldots$, an orthonormal basis of the vector space
$T_{h_{id}}(\Gamma_D)$ under some metric, e.g., the basis presented in~\cite{riseth} which we
describe later in this section,
at the $(k+1)^{th}$ iteration of the approach the gradient of $H_{k+1}$ is approximated
by $\nabla H_{k+1} = \sum_{i=1}^N (\nabla_{w_i} H_{k+1}) w_i$, for $N$ large enough.
If under the same metric $\nabla H_{k+1}$ is considered to be small enough, then $h^{(k)}$
is taken to be the solution of the gradient descent approach, although perhaps a local solution.
Otherwise, using a small step size $\delta >0$, $h_{k+1}$ and $h^{(k+1)}$
are computed, $h_{k+1}=h_{id}-\delta\;\nabla H_{k+1}$,
$h^{(k+1)}=h^{(k)}\circ h_{k+1}$. That $h^{(k+1)}$ is computed so that
$H(h^{(k+1)})=H_{k+1}(h_{k+1})$ follows by letting $\tilde{q}_2$ be as obtained at the
$k^{th}$ iteration of the approach, and proving that
\[ q_2(h^{(k+1)}(r,t))\sqrt{\frac{\partial h^{(k+1)}}{\partial (r,t)}(r,t)}
=\tilde{q}_2(h_{k+1}(r,t))\sqrt{\frac{\partial h_{k+1}}{\partial (r,t)}(r,t)},\; (r,t)\in D, \]
as follows:
\[ \tilde{q}_2(h_{k+1}(r,t))\sqrt{\frac{\partial h_{k+1}}{\partial (r,t)}(r,t)} \]
\begin{eqnarray*}
	&=& q_2(h^{(k)}(h_{k+1}(r,t)))\sqrt{\frac{\partial h^{(k)}}{\partial (u,w)}(h_{k+1}(r,t))}
	\sqrt{\frac{\partial h_{k+1}}{\partial (r,t)}(r,t)}\\
	&=&q_2(h^{(k)}(h_{k+1}(r,t)))\sqrt{\frac{\partial h^{(k)}}{\partial (u,w)}(h_{k+1}(r,t))
	{\frac{\partial h_{k+1}}{\partial (r,t)}(r,t)}}\\
	&=&q_2(h^{(k+1)}(r,t))\sqrt{\frac{\partial h^{(k+1)}}{\partial (r,t)}(r,t)}
\end{eqnarray*}
with $(u(r,t),w(r,t)) = h_{k+1}(r,t)$
by the product rule for determinants and the chain rule.
\\ \smallskip \\
The lemma and theorem follow. Here shape functions are $C^1$ as well.
\\ \smallskip\\
As above, given $h\in\Gamma_D$, in what follows 
$\frac{\partial h}{\partial (r,t)}$ is the determinant of the Jacobian of~$h$, i.e.,
$\frac{\partial h}{\partial (r,t)} =
\frac{\partial h_1}{\partial r}\frac{\partial h_2}{\partial t}
-\frac{\partial h_1}{\partial t}\frac{\partial h_2}{\partial r}$,
where $h(r,t)=(h_1(r,t),h_2(r,t))$, $(r,t)\in D$. Finally,
given $h\in\Gamma_D$, again $h(r,t)=(h_1(r,t),h_2(r,t))$, and $v\in T_h(\Gamma_D)$,
$v(r,t)=(v_1(r,t),v_2(r,t))$, $(r,t)\in D$, we define
\[ \frac{\partial (v,h)}{\partial (r,t)}\equiv
\frac{\partial v_1}{\partial r}\frac{\partial h_2}{\partial t}
-\frac{\partial v_2}{\partial r}\frac{\partial h_1}{\partial t}
+\frac{\partial h_1}{\partial r}\frac{\partial v_2}{\partial t}
-\frac{\partial h_2}{\partial r}\frac{\partial v_1}{\partial t}, \]
and let $\mathrm{div}(v)$ be the divergence of $v$, i.e., $\mathrm{div}(v)(r,t)=
\frac{\partial v_1}{\partial r}(r,t)+\frac{\partial v_2}{\partial t}(r,t)$.
\\ \smallskip \\
{\bf Lemma 2:} Given $q$, the shape function of a surface in $3-$dimensional space,
$\phi:\Gamma_D\rightarrow [q]_{\Gamma_D}$, $\phi(h(r,t))=
q(h(r,t))\sqrt{\frac{\partial h}{\partial (r,t)}(r,t)}$, then given $h$ in $\Gamma_D$, $v$ in
$T_{h}(\Gamma_D)$, with $\dot{q}$ the Jacobian of $q$, the differential of $\phi$ at $h$ applied on $v$ is
\begin{eqnarray*}
	(d\phi_h(v))(r,t)&=&\sqrt{\frac{\partial h}{\partial (r,t)}(r,t)}\dot{q}(h(r,t))v(r,t)\;+\\
	&\ & \frac{1}{2\sqrt{\frac{\partial h}{\partial (r,t)}(r,t)}}\frac{\partial (v,h)}{\partial (r,t)}(r,t)
	q(h(r,t)),\ (r,t)\in D.
\end{eqnarray*}
{\bf Proof:} Let $\alpha(\tau,\cdot,\cdot)$ be a differentiable path in $\Gamma_D$ passing through $h$ at
$\tau=0$, i.e., $\alpha(0,r,t) = h(r,t)$, $(r,t)\in D$. Let the velocity of this path at $\tau=0$ be given
by $v\in T_{h}(\Gamma_D)$, i.e., $v(r,t) = \frac{\partial\alpha}{\partial\tau}(0,r,t)$, $(r,t)\in D$. Note
as well, with $\alpha(\tau,r,t)=(\alpha_1(\tau,r,t),\alpha_2(\tau,r,t))$, and
again with $h(r,t)=(h_1(r,t),h_2(r,t))$, $v(r,t)=(v_1(r,t),v_2(r,t))$, that
\begin{eqnarray*}
	&\ & \frac{\partial\alpha_1}{\partial r}(0,r,t) = \frac{\partial h_1}{\partial r}(r,t),
\ \ \ \frac{\partial\alpha_1}{\partial t}(0,r,t) = \frac{\partial h_1}{\partial t}(r,t),\\
	&\ &\frac{\partial\alpha_2}{\partial r}(0,r,t) = \frac{\partial h_2}{\partial r}(r,t),
\ \ \ \frac{\partial\alpha_2}{\partial t}(0,r,t) = \frac{\partial h_2}{\partial t}(r,t),\\
\end{eqnarray*}
\begin{eqnarray*}
	&\ & \frac{\partial^2\alpha_1}{\partial r\partial\tau}(0,r,t) = \frac{\partial v_1}{\partial r}(r,t),
\ \ \ \frac{\partial^2\alpha_1}{\partial t\partial\tau}(0,r,t) = \frac{\partial v_1}{\partial t}(r,t),\\
	&\ &\frac{\partial^2\alpha_2}{\partial r\partial\tau}(0,r,t) = \frac{\partial v_2}{\partial r}(r,t),
\ \ \ \frac{\partial^2\alpha_2}{\partial t\partial\tau}(0,r,t) = \frac{\partial v_2}{\partial t}(r,t),
\ \ (r,t)\in~D.
\end{eqnarray*}
Since
\begin{eqnarray*}
	\frac{\partial\alpha}{\partial (r,t)}(\tau,r,t)&=&
	\frac{\partial\alpha_1}{\partial r}(\tau,r,t)
	 \frac{\partial\alpha_2}{\partial t}(\tau,r,t)
	-\frac{\partial\alpha_2}{\partial r}(\tau,r,t)
	 \frac{\partial\alpha_1}{\partial t}(\tau,r,t),
\end{eqnarray*}
it follows then that
\begin{eqnarray*}
	\frac{\partial\alpha}{\partial (r,t)}(\tau,r,t)|_{\tau=0}&=&
	\frac{\partial\alpha_1}{\partial r}(0,r,t)
	 \frac{\partial\alpha_2}{\partial t}(0,r,t)
	-\frac{\partial\alpha_2}{\partial r}(0,r,t)
	 \frac{\partial\alpha_1}{\partial t}(0,r,t)\\
	&=&\frac{\partial h_1}{\partial r}(r,t)
	 \frac{\partial h_2}{\partial t}(r,t)
	-\frac{\partial h_2}{\partial r}(r,t)
	 \frac{\partial h_1}{\partial t}(r,t)\\
	 &=&\frac{\partial h}{\partial (r,t)}(r,t).
\end{eqnarray*}
In addition, we note that
\begin{eqnarray*}
	&\ &\frac{\partial}{\partial\tau}\big(\frac{\partial\alpha}
	{\partial (r,t)}(\tau,r,t)\big)|_{\tau=0}\\
	&=&\frac{\partial}{\partial\tau}\big(\frac{\partial\alpha_1}{\partial r}(\tau,r,t)
	 \frac{\partial\alpha_2}{\partial t}(\tau,r,t)\;-
	 \frac{\partial\alpha_2}{\partial r}(\tau,r,t)
	 \frac{\partial\alpha_1}{\partial t}(\tau,r,t)\big)|_{\tau=0}\\
	&=&\big(\frac{\partial^2\alpha_1}{\partial\tau r}(\tau,r,t)
	 \frac{\partial\alpha_2}{\partial t}(\tau,r,t)+
	 \frac{\partial\alpha_1}{\partial r}(\tau,r,t)
		\frac{\partial^2\alpha_2}{\partial\tau t}(\tau,r,t)\\
	& &-\frac{\partial^2\alpha_2}{\partial\tau r}(\tau,r,t)
	 \frac{\partial\alpha_1}{\partial t}(\tau,r,t)-
	 \frac{\partial\alpha_2}{\partial r}(\tau,r,t)
		\frac{\partial^2\alpha_1}{\partial\tau t}(\tau,r,t)\big)|_{\tau=0}\\
	&=&\big(\frac{\partial^2\alpha_1}{\partial r \tau}(\tau,r,t)
	 \frac{\partial\alpha_2}{\partial t}(\tau,r,t)+
	 \frac{\partial\alpha_1}{\partial r}(\tau,r,t)
		\frac{\partial^2\alpha_2}{\partial t \tau}(\tau,r,t)\\
	& &-\frac{\partial^2\alpha_2}{\partial r \tau}(\tau,r,t)
	 \frac{\partial\alpha_1}{\partial t}(\tau,r,t)-
	 \frac{\partial\alpha_2}{\partial r}(\tau,r,t)
		\frac{\partial^2\alpha_1}{\partial t \tau}(\tau,r,t)\big)|_{\tau=0}\\
	&=&\frac{\partial v_1}{\partial r}(r,t)
	 \frac{\partial h_2}{\partial t}(r,t)+
	 \frac{\partial h_1}{\partial r}(r,t)
		\frac{\partial v_2}{\partial t}(r,t)\\
	& &-\frac{\partial v_2}{\partial r}(r,t)
	 \frac{\partial h_1}{\partial t}(r,t)-
	 \frac{\partial h_2}{\partial r}(r,t)
	 \frac{\partial v_1}{\partial t}(r,t)\\
	&=&\frac{\partial (v,h)}{\partial (r,t)}(r,t).
\end{eqnarray*}
Since $\alpha(\tau,\cdot,\cdot)$ is a path in $\Gamma_D$, then $\phi(\alpha(\tau,\cdot,\cdot))$ is the
corresponding path in $[q]_{\Gamma_D}$, and since $v$ is the velocity of $\alpha(\tau,\cdot,\cdot)$ at
$\tau=0$, then the velocity of $\phi(\alpha(\tau,\cdot,\cdot))$ at $\tau=0$ is $d\phi_h(v)$,\\
\smallskip\\
that is to say
\[ (d\phi_h(v))(r,t)=\frac{\partial}{\partial\tau}|_{\tau=0}\;\phi(\alpha(\tau,r,t)) \]
\begin{eqnarray*}
	&=&\frac{\partial}{\partial\tau}|_{\tau=0}\;\Big(\sqrt{\frac{\partial\alpha}{\partial (r,t)}(\tau,r,t)}
	\;q(\alpha(\tau,r,t))\Big)\\
	&=&\Big(\sqrt{\frac{\partial\alpha}{\partial (r,t)}(\tau,r,t)}
	\;\dot{q}(\alpha(\tau,r,t))\;\frac{\partial\alpha}{\partial\tau}(\tau,r,t)\;+\\
	&\ & \frac{1}{2\sqrt{\frac{\partial\alpha}{\partial (r,t)}(\tau,r,t)}}
	\;\frac{\partial}{\partial \tau}\big(\frac{\partial\alpha}{\partial (r,t)}(\tau,r,t)\big)
	\;q(\alpha(\tau,r,t))\Big)|_{\tau=0}\\
	&=&\sqrt{\frac{\partial h}{\partial (r,t)}(r,t)}\dot{q}(h(r,t))v(r,t)\;+\\
	&\ & \frac{1}{2\sqrt{\frac{\partial h}{\partial (r,t)}(r,t)}}\frac{\partial (v,h)}
	{\partial (r,t)}(r,t) q(h(r,t)),\ (r,t)\in D.
\end{eqnarray*}
\endproof\\
{\bf Corollary 2:} Given $q$, $\phi$, $v$ as above, then
\[ (d\phi_{h_{id}}(v))(r,t)=\dot{q}(r,t)v(r,t)+ \frac{1}{2}\mathrm{div}(v)(r,t)q(r,t),\ (r,t)\in D.\]
{\bf Theorem 2:}
With $H_{k+1}(h) = \int_0^1\int_0^1 ||q_1(r,t)
-\tilde{q}_2(h(r,t))\sqrt{\frac{\partial h}{\partial (r,t)}(r,t)}||^2\;dr\;dt$
so that $\tilde{q}_2(r,t)=q_2(h^{(k)}(r,t))\sqrt{\frac{\partial h^{(k)}}{\partial (r,t)}(r,t)}$
from the definition of $H_{k+1}$,
then the directional derivative of $H_{k+1}$ in the direction of $v\in T_{h_{id}}(\Gamma_D)$ is
\begin{eqnarray*}
	\nabla_v H_{k+1} &=& -2\int_0^1\int_0^1\Big\langle q_1(r,t)-\tilde{q}_2(r,t)\;,\;\\
	&\ &\dot{\tilde{q}}_2(r,t)v(r,t) +\frac{1}{2}
	\mathrm{div}(v)(r,t) \tilde{q}_2(r,t)
	\Big\rangle\;dr\;dt.
\end{eqnarray*}
{\bf Proof:} Let $\alpha(\tau,\cdot,\cdot)$ be a differentiable path in $\Gamma_D$ passing through
$h_{id}$ at $\tau=0$, i.e., $\alpha(0,r,t)=h_{id}(r,t)$, $(r,t)\in D$, with the velocity of this
path at $\tau=0$ equal to $v$, i.e., $\frac{\partial\alpha}{\partial\tau}(0,r,t)=v(r,t)$, $(r,t)\in D$.
\\ \smallskip\\
Note,
\[ H_{k+1}(\alpha(\tau,r,t))=\int_0^1\int_0^1 \Big\langle q_1(r,t)
	-\tilde{q}_2(\alpha(\tau,r,t))\sqrt{\frac{\partial \alpha}{\partial (r,t)}(\tau,r,t)}\;,\; \]
\begin{eqnarray*}
	&\ &q_1(r,t) -\tilde{q}_2(\alpha(\tau,r,t))\sqrt{\frac{\partial \alpha}{\partial (r,t)}
	(\tau,r,t)}\Big\rangle\;dr\;dt.
\end{eqnarray*}
Again, since in general
$\frac{d}{ds}\langle f(s),g(s)\rangle = \langle f(s),g'(s)\rangle+\langle f'(s),g(s)\rangle$, from
which $\frac{d}{ds}\langle f(s),f(s)\rangle = 2\langle f(s),f'(s)\rangle$,
then differentiating $H_{k+1}(\alpha(\tau,r,t))$ with respect to $\tau$ by differentiating this last
integral with respect to $\tau$ (done under the integral), and setting $\tau$ equal to zero, gives
\begin{eqnarray*}
	\nabla_v H_{k+1} &=& 2\int_0^1\int_0^1 \Big\langle q_1(r,t)
	-\tilde{q}_2(h_{id}(r,t))\sqrt{\frac{\partial h_{id}}{\partial (r,t)}(r,t)}\;,\;\\
	&\ & \frac{\partial}{\partial\tau}|_{\tau=0}
	\Big(-\tilde{q}_2(\alpha(\tau,r,t))\sqrt{\frac{\partial \alpha}{\partial (r,t)}
	(\tau,r,t)}\Big)\Big\rangle\;dr\;dt\\
	&=& -2\int_0^1\int_0^1 \Big\langle q_1(r,t) -\tilde{q}_2(r,t)\;,\;\\
	&\ & \frac{\partial}{\partial\tau}|_{\tau=0}
	\Big(\tilde{q}_2(\alpha(\tau,r,t))\sqrt{\frac{\partial \alpha}{\partial (r,t)}
	(\tau,r,t)}\Big)\Big\rangle\;dr\;dt\\
	&=& -2\int_0^1\int_0^1 \Big\langle q_1(r,t) -\tilde{q}_2(r,t)\;,\;\\
	&\ &\dot{\tilde{q}}_2(r,t)v(r,t) +\frac{1}{2}
	\mathrm{div}(v)(r,t) \tilde{q}_2(r,t) \Big\rangle\;dr\;dt
\end{eqnarray*}
by Corollary 2 and the proof of Lemma 2 as $\alpha$ here is the same as $\alpha$
there with $h$ equal to~$h_{id}$. \endproof
\\ \smallskip\\
Note that given $h\in\Gamma_D$ and a differentiable path $\alpha(\tau,\cdot,\cdot)$ in $\Gamma_D$
through $h$ at $\tau=0$, i.e., $\alpha(0,r,t) = h(r,t)$, $(r,t)\in D$,
then with $\alpha(\tau,r,t)=(\alpha_1(\tau,r,t),\alpha_2(\tau,r,t))$,
since for any real numbers $\tau_1$, $\tau_2$, close or equal to 0, for $0\leq r,t \leq 1$,
$\alpha_1(\tau_1,0,t) - \alpha_1(\tau_2,0,t) = 0-0=0$, 
$\alpha_2(\tau_1,r,0) - \alpha_2(\tau_2,r,0) = 0-0=0$, 
$\alpha_1(\tau_1,1,t) - \alpha_1(\tau_2,1,t) = 1-1=0$, 
$\alpha_2(\tau_1,r,1) - \alpha_2(\tau_2,r,1) = 1-1=0$, 
with $v(r,t)=(v_1(r,t),v_2(r,t))$, $(r,t)\in D$, $v\in T_h(\Gamma_D)$,
it can also be shown that
\begin{eqnarray*}
	T_{h}(\Gamma_D)&=&\{v:D\rightarrow \mathbb{R}^2\;|\;
	v_1(0,t)=v_2(r,0)=v_1(1,t)=v_2(r,1)=0,\\
	&\ &\ \; 0\leq r,t\leq 1,\;v\ \mathrm{smooth}\}.
\end{eqnarray*}
Next, we present and describe the orthonormal basis under the $L^2$ norm of the
vector space $T_{h_{id}}(\Gamma_D)$ presented in \cite{riseth} as it is the basis we use as well.
As described in \cite{riseth}, first an orthonormal basis $B^1$ under the $L^2$ norm is identified
for the space
\[ S_{rt}^r=\{v:D\rightarrow \mathbb{R}\;|\; v(0,t)=v(1,t)=0,\; 0\leq t\leq 1,\;
v\ \mathrm{smooth}\} \]
that consists of three families of functions: $\sqrt{2} \sin(\pi kr)$, $2 \sin(\pi kr) \cos(2\pi lt)$,
$2 \sin(\pi kr) \sin(2\pi lt)$, $k,l =1,2,3,\ldots$, $(r,t)\in D$.
\\ \smallskip \\
With one$(t) = 1$ for all $t,\; 0\leq t\leq 1$,
using $\phi_m(r)$, $m=1,2,3,\ldots$, to refer to $\sqrt{2} \sin(\pi kr)$,
$0\leq r\leq 1$, $k=1,2,3,\ldots$,
and $\psi_n(t)$, $n=1,2,3,\ldots$, to refer to one$(t)$, $\sqrt{2} \cos(2\pi lt)$,
$\sqrt{2} \sin(2\pi lt)$, $0\leq t\leq 1$, $l=1,2,3,\ldots$,
then under the $L^2$ norm $\phi_m$, $m=1,2,3,\ldots$, is an orthonormal basis for the space
\[ \{v:[0,1]\rightarrow \mathbb{R}\;|\; v(0)=v(1)=0,\; v\ \mathrm{smooth}\} \]
and $\psi_n$, $n=1,2,3,\ldots$, is an orthonormal basis for the space
\[ \{v:[0,1]\rightarrow \mathbb{R}\;|\; v\ \mathrm{smooth}\}. \]
Clearly $B^1$ is the tensor product of the two bases $\phi_m$, $m=1,2,3\ldots$, and $\psi_n$,
$n=1,2,3,\ldots$, and that $B_1$ is linearly independent is a direct result of the linear
independence of these two bases. Finally, given $v\in S_{rt}^r$, that $v$ is a linear combination,
possibly infinite under the $L^2$ norm, of elements of $B_1$, is established by applications of
Parseval's identity (three times) together with Fubini's theorem (twice) as~follows:
	\[	\sum_{m,n} \Big|\int\int_D v(r,t)\phi_m(r)\psi_n(t)drdt\Big|^2
	 =  \sum_{m,n} \Big|\int_0^1\Big(\int_0^1 v(r,t)\phi_m(r)dr\Big)\psi_n(t)dt\Big|^2 \]
\begin{eqnarray*}
	&=& \sum_m \int_0^1\Big|\int_0^1 v(r,t)\phi_m(r)dr\Big|^2dt
	= \int_0^1 \sum_m\Big|\int_0^1 v(r,t)\phi_m(r)dr\Big|^2dt\\
	&=& \int_0^1\Big(\int_0^1\Big| v(r,t)\Big|^2dr\Big) dt
	= \int\int_D\Big| v(r,t)\Big|^2drdt.
\end{eqnarray*}
Similarly, an orthonormal basis $B^2$ under the $L^2$ norm can be identified and justified
for the space
\[ S_{rt}^t=\{v:D\rightarrow \mathbb{R}\;|\; v(r,0)=v(r,1)=0,\; 0\leq r\leq 1,\;
v\ \mathrm{smooth}\} \]
that consists of three families of functions: $\sqrt{2} \sin(\pi kt)$, $2 \sin(\pi kt) \cos(2\pi lr)$,
$2 \sin(\pi kt) \sin(2\pi lr)$, $k,l =1,2,3,\ldots$, $(r,t)\in D$.
\\ \smallskip\\
Using $\eta_j(r,t)$, $j=1,2,3,\ldots$, to refer to the elements of the basis $B^1$ of $S_{rt}^r$,
i.e., to $\sqrt{2} \sin(\pi kr)\mathrm{one}(t)$, $2 \sin(\pi kr) \cos(2\pi lt)$,
$2 \sin(\pi kr) \sin(2\pi lt)$, $k,l =1,2,3,\ldots$, $(r,t)\in D$, it is clear then that
$\eta_j(t,r)$, $j=1,2,3,\ldots$, $(r,t)\in D$, are the elements of the basis $B^2$ of $S_{rt}^t$.
Note as well that if for some integer $KL>0$ we restrict $k,l$ above to range from $1$ to $KL$, then
$j$ above in the definition of $\eta_j$ will range from $1$ to $KL+2(KL)^2$.
\\ \smallskip\\
With zero$(r,t)=0$ for all $(r,t)\in D$, then as in \cite{riseth} an orthonormal basis $B$ of
$T_{h_{id}}(\Gamma_D)$ under the $L^2$ norm can be identified:
\[ B=\{(\eta_j(r,t),\mathrm{zero}(r,t)),(\mathrm{zero}(r,t),\eta_j(t,r)),j=1,2,3,\ldots, (r,t)\in D\}. \]
We note that under the identification of $T_h(\Gamma_D)$ given above following the
proof of Theorem~2, each element of $B$ is indeed in~$T_{h_{id}}(\Gamma_D)$. We also note that with
$KL$ as above so that $k,l$ above are restricted to range from $1$ to $KL$, and $j$ above in the definition
of $\eta_j$ is therefore restricted to range from $1$ to $KL+2(KL)^2$, if in the definition above of the
basis $B$, $j$ is restricted as well to range from $1$ to $KL+2(KL)^2$, then $B$ is truncated to have
$2(KL+2(KL)^2)$ elements.
\\ \smallskip\\
Next, we present a simplified outline of the procedure for the elastic shape registration of two
surfaces in $3-$dimensional space using gradient descent and dynamic programming. Note that
in the procedure a step size $\delta >0$ is used for computing elements of $\Gamma_D$, the
computation of $\delta$ discussed below, $\delta$ as large as the gradient descent approach
allows but small enough to guarantee that each computed element of $\Gamma_D$ is indeed
in $\Gamma_D$, in particular that the determinant of its Jacobian is positive everywhere on~$D$.
We note as well that in this simplified outline of the procedure, rotations are also taken into
account as they should be, so that it is \[ E(h,R)=\int_0^1\int_0^1 \big|\big|Rq_1
-(q_2\circ h)\sqrt{\frac{\partial h}{\partial (r,t)}}\big|\big|^2dr\,dt \] that we actually hope
to minimize with respect to $h\in \Gamma_D$ and $R\in SO(3)$.
Here we use $w_i$, $i=1,2,3,\ldots$, to refer to the elements of the basis~$B$.
\\ \smallskip\\
{\bf Simplified Outline of Optimization Procedure}\\
1. With $q_1$, $q_2$ as the shape functions of the two simple surfaces under consideration, say $S_1$
and $S_2$, execute dynamic-programming-based Procedure DP-surface-min in \cite{bernal5} for $q_1$,
$q_2$, to obtain (partially) optimal $\hat{R}\in SO(3)$ and $h_0\in\Gamma_D$,
where $\hat{R}$ rotates $S_1$ and $h_0$ reparametrizes~$S_2$.\\
Set $k=0$, $h^{(0)}=h_0$.\\
Compute $\hat{q}_1=\hat{R}q_1$,
$\tilde{q}_2= (q_2,h^{(0)})=(q_2\circ h^{(0)})\sqrt{\frac{\partial h^{(0)}}{\partial (r,t)}}$,\\
and $E(h^{(0)},\hat{R})=\int_0^1\int_0^1 ||\hat{q}_1-\tilde{q}_2||^2dr\,dt$.\\
If $E(h^{(0)},\hat{R})$ is equal or close to zero, then go to Step~4 below.\\
Set $N$ to a positive integer large enough.\\
2. For each $i$, $i=1,\ldots,N$, compute $\nabla_{w_i} H_{k+1} =\\
-2\int_0^1\int_0^1\Big\langle \hat{q}_1(r,t)-\tilde{q}_2(r,t)\;,\;
\dot{\tilde{q}}_2(r,t)w_i(r,t) +\frac{1}{2}
\mathrm{div}(w_i)(r,t) \tilde{q}_2(r,t) \Big\rangle\;dr\;dt$\\
as indicated by Theorem~2 above, and compute
$\nabla H_{k+1} = \sum_{i=1}^N (\nabla_{w_i} H_{k+1}) w_i$.\\
If the $L^2$ norm of $\nabla H_{k+1}$ is small enough, then go to Step~3 below.\\
Else for $\delta >0$ appropriately chosen, set $h_{k+1}=h_{id}-\delta\,\nabla H_{k+1}$,\\
$h^{(k+1)}=h^{(k)}\circ h_{k+1}$, and $k=k+1$.\\
Compute
$\tilde{q}_2= (q_2,h^{(k)})=(q_2\circ h^{(k)})\sqrt{\frac{\partial h^{(k)}}{\partial (r,t)}}$,\\
and $E(h^{(k)},\hat{R})=\int_0^1\int_0^1 ||\hat{q}_1-\tilde{q}_2||^2dr\,dt$.\\
If $E(h^{(k)},\hat{R})$ is equal or close to zero, then set $h_0=h^{(k)}$ and  go to Step~4 below.\\
If $E(h^{(k)},\hat{R})$ is not much less than $E(h^{(k-1)},\hat{R})$,
then go to Step~3~below.\\
Else repeat this step (Step 2).\\
3. Set $h_0=h^{(k)}$, $E_L=E(h^{(k)},\hat{R})$.\\
With $S_1$, $S_2$ as above, execute KU3 algorithm in \cite{bernal5}, that is, the Kabsch-Umeyama
algorithm \cite{kabsch1,kabsch2,lawrence,umeyama}, for $\tilde{q}_2$, $q_1$, to obtain optimal
$\hat{R}\in SO(3)$ for the rigid alignment of $S_1$ and $\tilde{S}_2$, where $\hat{R}$ rotates $S_1$, and
$\tilde{S}_2$ is $S_2$ reparametrized so that $\tilde{q}_2$ is its shape function.\\
Compute $\hat{q}_1 = \hat{R}q_1$, and
$E(h_0,\hat{R})=\int_0^1\int_0^1 ||\hat{q}_1-\tilde{q}_2||^2dr\,dt$.\\
If $E(h_0,\hat{R})$ is equal or close to zero, then go to Step~4 below.\\
If this step (Step 3) has been executed enough times or
$E(h_0,\hat{R})$ is not much less than $E_L$, then go to Step~4~below.\\
Else set $k=0$, $h^{(0)}=h_0$.\\
Go to Step 2 above.\\
4. $h=h_0$ and $R=\hat{R}$ minimize $E(h,R)$, possibly resulting in a local solution.\\
If not a local solution, then $E(h_0,\hat{R})^{1/2}$ is the elastic shape distance between
the two surfaces.\\
Stop.
\\ \smallskip\\
Finally, we discuss how the step size $\delta>0$ mentioned above for computing elements of
$\Gamma_D$ is chosen. This is done in a way similar to what is done in~\cite{riseth}.
In particular, in Step~2 of the simplified outline of the optimization procedure above,
$h_{k+1}$ is computed as $h_{id}-\delta\,\nabla H_{k+1}$ so that $\delta>0$ should be as large
as the gradient descent approach allows but small enough that the determinant of the Jacobian
of $h_{k+1}$ is positive at every $(r,t)$ in~$D$, i.e.,
$\frac{\partial h_{k+1}}{\partial (r,t)}(r,t)>0$ for each $(r,t)\in D$.
\\ \smallskip\\
Let $A(r,t)$ be the $2\times 2$ matrix which is the Jacobian of $\nabla H_{k+1}(r,t)$,
$(r,t)\in D$. It then follows that the Jacobian of
$h_{k+1}(r,t)=(h_{id}-\delta\,\nabla H_{k+1})(r,t)$, $(r,t)\in D$, is a $2\times 2$ matrix
as well equal to $I_2-\delta\,A(r,t)$, where $I_2$ is the $2\times 2$ identity matrix. Thus,
it is not hard to show that 
\[ \frac{\partial h_{k+1}}{\partial (r,t)}(r,t)=
1-\mathrm{tr}(A(r,t))\,\delta+\mathrm{det}(A(r,t))\,\delta^2,\ \ (r,t)\in D, \]
where tr stands for trace and det for determinant.
\\ \smallskip\\
For each $(r,t)\in D$, we hope to compute $\delta(r,t)$, which is the largest positive number
for which $\frac{\partial h_{k+1}}{\partial (r,t)}(r,t)>0$ for $\delta$ in the interval
$(0,\delta(r,t))$, $\delta(r,t)$ possibly equal to~$\infty$. With $a=\mathrm{det}A(r,t)$,
$b=-\mathrm{tr}A(r,t)$, $c=1$, if $a$ is zero or close to zero, then it is not hard to show
that $\delta(r,t)$ is approximately $-1/b$ if $b<0$, $\infty$ otherwise. On the other hand,
if $a$ is not close to zero, then applying the quadratic formula it is not hard to show that
$\delta(r,t)$ equals $\frac{-b-\sqrt{b^2-4a}}{2a}$ if this last number is a positive real
number, $\infty$ otherwise.
\\ \smallskip\\
Ideally we would like to be able to compute $\delta_{min}=\mathrm{min}_{(r,t)\in D}
\,\delta(r,t)$, and if this number is positive, identify as the desired step size $\delta$
a positive number slightly less than $\delta_{min}$ so that 
$\frac{\partial h_{k+1}}{\partial (r,t)}(r,t)>0$ for every $(r,t)\in D$. However, since as
pointed out in \cite{bernal5}, in practice we can only work with a discretization of $D$,
minimizing $\delta(r,t)$ over $D$ is a moot point, and in fact
we assume that positive integers $K$, $L$, not necessarily equal, and partitions of $[0,1]$,
$\{r_i\}_{i=1}^K$, $r_1=0<r_2<\ldots <r_K=1$, $\{t_j\}_{j=1}^L$, $t_1=0<t_2<\ldots <t_L=1$,
not necessarily uniform, are given, so that it is the grid $G$ on $D$,
$G=\{(r_i,t_j),i=1,\ldots,K, \,j=1,\ldots,L\}$ that is actually used instead of $D$ to
identify the desired step size~$\delta$, i.e., we compute
$\delta_{min}$ as $\mathrm{min}_{(r,t)\in G}\,\delta(r,t)$ instead, still identifying as the
desired step size $\delta$ a positive number slightly less than~$\delta_{min}$.
Assuming then for all intents and purposes that with this $\delta$ the determinant of the
Jacobian of $h_{k+1}$ is positive on $D$, we may assume as well that the determinant of
the Jacobian of $h^{(k+1)}$ is positive on $D$ by the product rule for determinants
and the chain rule.
\\ \smallskip\\
At the risk of making the step size $\delta$ too small, we may reduce the step size computed
as suggested above, to make sure that indeed $h_{k+1}$, and therefore $h^{(k+1)}$, is one to
one and maps $D$ onto~$D$, while maintaining the boundary conditions of elements of~$\Gamma_D$.
For this purpose we take advantage of the Gale-Nikaido Theorem~\cite{gale} that follows.
Here, given integer $n>0$, real numbers $a_i$, $b_i$, $i=1,\ldots,n$, some or all of them
allowed to be $-\infty$ or $\infty$, a rectangular region $R$ in $\mathbb{R}^n$ is defined by
\[ R=\{x: x\in\mathbb{R}^n,\ x=(x_1,\ldots,x_n) \mathrm{\ with\ } a_i\leq x_i\leq b_i,
\ i=1,\ldots,n\}. \]
{\bf Theorem 3:} (Gale-Nikaido Theorem) If $F$ is a $C^1$ mapping from a rectangular region $R$
in $\mathbb{R}^n$ into $\mathbb{R}^n$ such that for all $x\in R$ each principle minor of the
Jacobian matrix of $F$ at $x$ is positive, then $F$ is injective.
\\ \smallskip\\
Accordingly, for each $(r,t)\in D$, with $a_{ij}(r,t)$, $i,j=1,2$, the entries of $A(r,t)$,
$A(r,t)$ as above, it is not hard to see that for our purposes the minors of interest of the
Jacobian matrix of $h_{k+1}$ at $(r,t)$ are $\frac{\partial h_{k+1}}{\partial (r,t)}(r,t)$,
$1-\delta\,a_{11}(r,t)$ and $1-\delta\,a_{22}(r,t)$, and the goal is then to compute
$\hat{\delta}(r,t)$, which is the largest positive number for which all three minors are positive
for $\delta$ in the interval $(0,\hat{\delta}(r,t))$, $\hat{\delta}(r,t)$ possibly equal to~$\infty$.
Again with $G$ as above, we work with $G$ instead of $D$, and note that for $(r,t)\in G$,
$\frac{\partial h_{k+1}}{\partial (r,t)}(r,t)$ has already been taken care of above during the
computation of $\delta_{min}$. Thus, for each $(r,t)\in G$, we compute $\hat{\delta}(r,t)$
only with respect to $1-\delta\,a_{11}(r,t)$ and $1-\delta\,a_{22}(r,t)$ as follows.
If both $a_{11}(r,t)$ and $a_{22}(r,t)$ are nonpositive, then $\hat{\delta}(r,t)$
equals~$\infty$. If both $a_{11}(r,t)$ and $a_{22}(r,t)$ are positive, then
$\hat{\delta}(r,t)$ is the smaller of $1/a_{11}(r,t)$ and $1/a_{22}(r,t)$. Otherwise,
only one of $a_{11}(r,t)$ and $a_{22}(r,t)$ is positive, and $\hat{\delta}(r,t)$ is 
1~divided by the one of the two that is positive. Having done this for each $(r,t)\in G$,
with $\delta_{min}$ as computed above, and
$\hat{\delta}_{min} = \mathrm{min}_{(r,t)\in G}\,\hat{\delta}(r,t)$,
we identify as the desired step size $\delta$ a positive number slightly less than the smaller
of $\delta_{min}$ and $\hat{\delta}_{min}$, and assume for all intents and purposes that
with this~$\delta$ the determinant of the Jacobian of $h_{k+1}$ is positive on~$D$, and since
$D$ is the unit square, thus a rectangular region, that $h_{k+1}$ is one to one on~$D$ by the
Gale-Nikaido Theorem above. In addition, since we assume
$\frac{\partial h_{k+1}}{\partial (r,t)}(r,t)>0$ on~$D$, thus nonzero, by the
inverse function theorem we may assume the inverse of $h_{k+1}$ is a $C^1$ function.  
\\ \smallskip\\
Assuming then that $h_{k+1}$ is a one-to-one $C^1$ function on all of~$D$ with a $C^1$
inverse, we show that $h_{k+1}$ maps $D$ onto~$D$. For this purpose we need the two
well-known results that follow. Here a homeomorphism is a one-to-one continuous function
from a topological space onto another that has a continuous inverse function, and a simply
connected domain is a path-connected domain where one can continuously shrink any simple
closed curve into a point while remaining in the domain. For two-dimensional regions, a
simply connected domain is one without holes in~it. The two results appeared in
\cite{bernal5}, the first result a standard result in the field of topology, the
proof of the second result presented in~\cite{bernal5} for the sake of completeness.
\\ \smallskip\\
{\bf Theorem 4:} If $X$ and $Y$ are homeomorphic topological spaces, then $X$ is simply
connected if and only if $Y$ is simply connected.\\ \smallskip\\
{\bf Theorem 5:} Given $E$, a compact simply connected subset of $\mathbb{R}^2$,
and $h:E\rightarrow \mathbb{R}^2$, a homeomorphism, then $h$ maps the boundary of $E$ to
exactly the boundary of~$h(E)$.
\\ \smallskip\\
From these two theorems it then follows that $h_{k+1}(D)$ is simply connected and that
$h_{k+1}$ maps the boundary of $D$ to exactly the boundary of~$h_{k+1}(D)$. Note, in
particular, the boundary of $h_{k+1}(D)$ is then contained in~$h_{k+1}(D)$.
\\ \smallskip\\
Note, from the definition of $h_{k+1}$, that
$h_{k+1}(0,0)=(0,0)$, $h_{k+1}(0,1)=(0,1)$, $h_{k+1}(1,1)=(1,1)$, $h_{k+1}(1,0)=(1,0)$.
In particular, given $r$, $0<r<1$, then again from the definition of $h_{k+1}$,
$h_{k+1}(r,1) = (r',1)$, $r'$ a number not necessarily between 0 and~1, so that
$h_{k+1}(r,1)$ is in the line $t=1$ which contains the line segment with endpoints
$(0,1)$, $(1,1)$, i.e., the top side of the unit square~$D$. However this line segment is
connected, so that its image under $h_{k+1}$ in the line $t=1$ is connected and thus
must contain the line segment, and since $h_{k+1}$ is one to one, this image is exactly
the line segment. Since the same is true for the other three sides of~$D$,
then it follows that the boundary conditions of elements of $\Gamma_D$ are satisfied by
$h_{k+1}$, and that $h_{k+1}$ actually maps the boundary of~$D$ onto
itself.  However, we already know that $h_{k+1}$ maps the boundary of $D$ to exactly the
boundary of~$h_{k+1}(D)$, thus the boundary of $D$ and the boundary of $h_{k+1}(D)$ are
exactly the same. Since $D$ and $h_{k+1}(D)$ are both simply connected,
then~$h_{k+1}(D)=D$.
\section{\large Results from Implementation of Methods}
\begin{figure}
\centering
\begin{tabular}{ccc}
\includegraphics[width=0.4\textwidth]{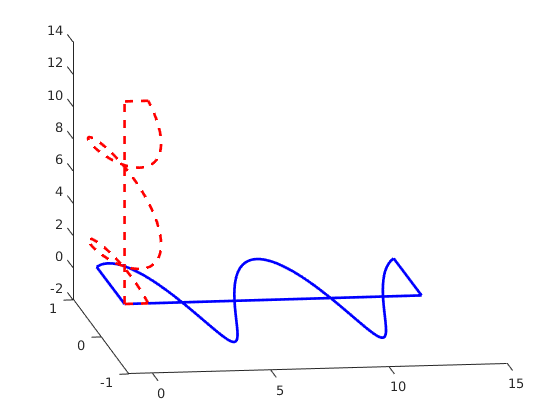}
\end{tabular}
\caption{\label{F:helicoids}
Boundaries of two surfaces of similar shape of the helicoid kind for $k=4$, type~$1$ in dashed red,
type~$2$ in solid blue.
}
\end{figure}

A software package has been implemented that incorporates the methods presented in the previous
section for computing, using gradient descent and dynamic programming, the elastic shape
registration of two simple surfaces in $3-$dimensional space, and therefore the elastic shape
distance between them. Actually, the software package consists of two separate pieces of software.
One piece is based on gradient descent as presented in the previous section for reparametrizing
one of the surfaces. This piece uses as the input initial solution the rotation and
reparametrization computed with the other piece of sofware in the package which is based on
dynamic programming as presented in \cite{bernal5} for reparametrizing one of the surfaces to
obtain a partial elastic shape registration of the surfaces. As described in \cite{bernal5},
the software in the package based on dynamic programming is in Matlab\footnote{The identification
of any commercial product or trade name does not imply endorsement or recommendation by the
National Institute of Standards and Technology.} with the exception of the dynamic programming
routine which is written in Fortran but is executed as a Matlab mex file. On the other hand, the
software in the package based on gradient descent is entirely in Matlab. In this section, we
present results obtained from executions of the software package. We note, the software package
as well as input data files, a README file, etc.  can be obtained at the following link\smallskip\\
%
\hspace*{.35in}\verb+https://doi.org/10.18434/mds2-3519+
\smallskip\\
We note, Matlab routine ESD\_\,main\_\,surf\_\,3d.m is the driver routine of the package, and Fortran
routine DP\_\,MEX\,\_\,WNDSTRP\_\,ALLDIM.F is the dynamic programming routine in the software based
on dynamic programming.  This routine has already been processed to be
executed as a Matlab mex file. In case the Fortran routine must be processed to obtain a new mex file,
this can be done by typing in the Matlab window:\\
mex -\,compatibleArrayDims DP\_\,MEX\,\_\,WNDSTRP\_\,ALLDIM.F
\\ \smallskip\\
As was the case for the software package described in \cite{bernal5}, at the start of the execution of
the new software package based on gradient descent and dynamic programming that we are describing here,
we assume $S_1$, $S_2$ are the two simple surfaces in $3-$dimensional space under consideration, with
functions $c_1$, $c_2: D\equiv [0,T_1]\times [0,T_2]\rightarrow \mathbb{R}^3$, $T_1$, $T_2>0$, as their
parametrizations, respectively, so that $S_1=c_1(D)$, $S_2=c_2(D)$. We also assume that as input to the
software, for positive integers $M$, $N$, not necessarily equal, and partitions of $[0,T_1]$, $[0,T_2]$,
respectively, $\{r_i\}_{i=1}^M$, $r_1=0<r_2<\ldots <r_M=T_1$, $\{t_j\}_{j=1}^N$,
$t_1=0<t_2<\ldots <t_N=T_2$, not necessarily uniform, discretizations of $c_1$, $c_2$ are given, each
discretization in the form of a list of $M\times N$ points in the corresponding surface, namely
$c_1(r_i,t_j)$ and $c_2(r_i,t_j)$, $i=1,\ldots,M$, $j=1,\ldots,N$, respectively, and for each $k$,
$k=1,2$, as specified in the Introduction section in \cite{bernal5}, in the order $c_k(r_1,t_1)$,
$c_k(r_2,t_1)$, $\ldots$, $c_k(r_M,t_1)$, $\ldots$, $c_k(r_1,t_N)$, $c_k(r_2,t_N)$, $\ldots$,
$c_k(r_M,t_N)$. Based on this input, for the purpose of computing, using dynamic programming, a partial
elastic shape registration of $S_1$ and $S_2$, together with the elastic shape distance between them
associated with the partial registration, the program always proceeds first to scale the partitions
$\{r_i\}_{i=1}^M$, $\{t_j\}_{j=1}^N$, so that they become partitions of $[0,1]$, and to compute an
approximation of the area of each surface. During the execution of the software package, the former
is accomplished by the driver routine Matlab routine 
ESD\_\,main\_\,surf\_\,3d.m, while the latter by Matlab routine ESD\_\,comp\_\,surf\_\,3d.m (called by
ESD\_\,main\_\,surf\_\,3d.m) through the computation for each $k$, $k=1,2$ of the sum of the areas of
triangles with vertices $c_k(r_i,t_j)$, $c_k(r_{i+1},t_{j+1})$, $c_k(r_i,t_{j+1})$, and $c_k(r_i,t_j)$,
$c_k(r_{i+1},t_j)$, $c_k(r_{i+1},t_{j+1})$, for $i=1,\ldots,M-1$, $j=1,\ldots,N-1$.
This last routine then proceeds to scale the discretizations of the parametrizations of the two
surfaces so that each surface has approximate area equal to~1 (given a surface and its approximate area,
each point in the discretization of the parametrization of the surface is divided by the square root of
half the approximate area of the surface).  Once routine ESD\_\,comp\_\,surf\_\,3d.m is done with these
computations, the actual computations of the rotation and reparametrization based on dynamic programming
for reparametrizing one of the surfaces to obtain a partial elastic shape registration of the surfaces,
are carried out by Matlab routine ESD\_\,core\_\,surf\_\,3d.m (called by ESD\_\,comp\_\,surf\_\,3d.m) in
which the methods for this purpose presented in \cite{bernal5}, mainly Procedure DP-surface-min in
Section~7 of \cite{bernal5}, have been implemented. Note, it is in this routine that the dynamic
programming routine Fortran routine DP\_\,MEX\,\_\,WNDSTRP\_\,ALLDIM.F is executed.
Finally, Matlab routine ESD\_\,grad\_\,surf\_\,3d.m (called by ESD\_\,main\_\,surf\_\,3d.m) is executed for
the purpose of computing, using gradient descent, the elastic shape registration of the two surfaces,
together with the elastic shape distance between them. Note, this routine uses as the input initial solution
the rotation and reparametrization computed by routine ESD\_\,core\_\,surf\_\,3d.m that as part of
the output of ESD\_\,comp\_\,surf\_\,3d.m become available for ESD\_\,grad\_\,surf\_\,3d.m to use as input.
Note as well that in this routine with integer $KL>0$ and infinite basis $B$, $KL$ and $B$ as defined in
the previous section, we use $KL=5$ so that the basis $B$ is then truncated to have $2(KL + 2(KL)^2)=110$
elements.
\\ \smallskip\\
The results that follow were obtained from applications of our software package on discretizations
of the three kinds of surfaces in $3-$dimensional space that were identified in \cite{bernal5} and that were
called there surfaces of the sine, helicoid and cosine-sine kind. Of course results in \cite{bernal5} were
obtained using software based on dynamic programming only on the aforementioned discretizations, while here
results were obtained through executions of our software package, using software based on both dynamic
programming and gradient descent, used separately and combined (by setting variable insol equal to~$1$ in
the driver routine Matlab routine ESD\_\,main\_\,surf\_\,3d.m, gradient descent is used with initial solution
the rotation and diffeomorphism computed with dynamic programming; otherwise, gradient descent is used with
initial solution the identity matrix and the identity diffeomorphism; note, with a couple of exceptions, the
results reported here obtained with the software package using dynamic programming before using gradient
descent are the same as the results reported in \cite{bernal5}, with comments given there about these
results still valid for the resuts here).  As was the case when using software based on
dynamic programming only as described in \cite{bernal5}, when using our software package, on input all
surfaces were given as discretizations on the unit square ($[0,1]\times [0,1]$), each interval $[0,1]$
uniformly partitioned into $100$ intervals so that the unit square was thus partitioned into $10000$
squares, each square of size $0.01\times 0.01$, their corners making up a set of~$10201$ points.
Using the same notation used above in this section, the uniform partitions of the
two $[0,1]$ intervals that define the unit square were then $\{r_i\}_{i=1}^M$, $\{t_j\}_{j=1}^N$, with
$M=N=101$, $r_{101}=t_{101}=1.0$, thus already scaled from the start as required, and by evaluating the
surfaces at the $10201$ points identified above in the order as specified above and in the Introduction
section in~\cite{bernal5}, a discretization of each surface was obtained consisting of~$10201$ points.
Given a pair of surfaces of one of the three kinds mentioned above, discretized as just described,
then during the execution of our software package on their discretizations, using software based on both
dynamic programming and gradient descent, used separately or combined,
one surface was identified as the first surface, the other one as the second surface (in the methods
presented in the previous section and in \cite{bernal5}, using gradient descent and/or dynamic programming,
the second surface is reparametrized while the first one is rotated). For the purpose of testing the
capability of the software, again using the same notation used above in this section, given $\gamma$, a
bijective function on the unit square to be defined below, with $(\hat{r}_i,\hat{t}_j)=\gamma(r_i,t_j)$,
$i=1,\ldots,101$, $j=1,\ldots,101$, the second surface was reparametrized through its discretization,
namely by setting $\hat{c}_2=c_2$ and computing $c_2(r_i,t_j) =\hat{c}_2(\hat{r}_i,\hat{t}_j)$,
$i=1,\ldots,101,\ j=1,\ldots,101$, while the first surface was kept as originally defined and discretized
by computing $c_1(r_i,t_j)$, $i=1,\ldots,101,\ j=1,\ldots,101$. All of the above done, the software
package then was executed twice, each time using the discretizations of the surfaces as just described in
terms of $c_1$, $c_2$, etc., to compute an approximation of the area of each surface and scale each surface
to have approximate area equal to~1, and then compute an elastic shape registration of the two surfaces and
the elastic shape distance between them associated with the registration, the first time using gradient
descent with dynamic programming, the second time using gradient descent without dynamic programming.
Note, in what follows, numbers obtained as elastic shape distances are actually the square of these
distances.
\\ \smallskip\\
The first results that follow were obtained from applications of our software package on discretizations of
surfaces in $3-$dimensional space that are actually graphs of functions based on the sine curve.
Given $k$, a positive integer, one type of surface to which we refer as a surface of the sine kind (type~1) is
defined by
\[ x(r,t)= r,\ \ \ y(r,t) = t,\ \ \ z(r,t) = \sin k \pi  r,\ \ \ (r,t) \in [0,1]\times [0,1], \]
and another one (type~2) by
\[ x(r,t) = \sin k \pi  r,\ \ \ y(r,t)= r,\ \ \ z(r,t) = t,\ \ \ (r,t) \in [0,1]\times [0,1], \]
the former a rotation of the latter by applying the rotation matrix
$\left( \begin{smallmatrix} 0 & 1 & 0\\ 0 & 0 & 1\\ 1 & 0 & 0\\ \end{smallmatrix} \right)$
on the latter, thus of similar shape.
\begin{figure}
\centering
\begin{tabular}{ccc}
\includegraphics[width=0.3\textwidth]{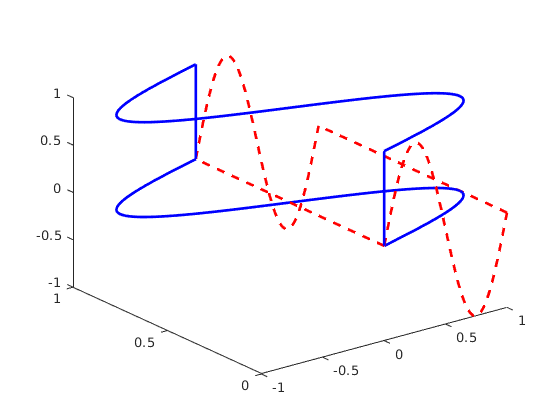}
&
\includegraphics[width=0.3\textwidth]{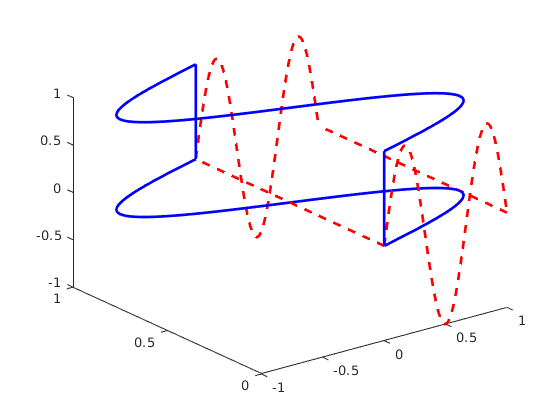}
&
\includegraphics[width=0.3\textwidth]{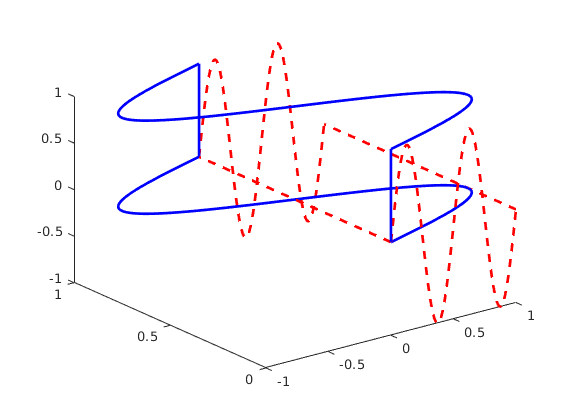}
\end{tabular}
\caption{\label{F:surfaces}
Three plots of boundaries of surfaces of the sine kind. Elastic shape registrations of the two
surfaces in each plot were computed using gradient descent, with and without dynamic programming.}

\end{figure}
\\ \smallskip\\
Three plots depicting surfaces (actually their boundaries in solid blue and dashed red) of the sine
kind for different values of $k$ are shown in Figure~\ref{F:surfaces}.
(Note that in the plots there, the $x-$, $y-$, $z-$ axes are not always to scale relative to one another).
In each plot two surfaces of the sine kind appear. The two surfaces in the leftmost plot being of similar
shape, clearly the elastic shape distance between them is exactly zero, and the hope was then that the
execution of our software package applied on these two surfaces, using gradient descent and/or dynamic
programming, would produce an elastic shape distance between them equal or close to zero. The type~2 surface
in each plot (in solid blue) was considered to be the first surface in the plot. In each plot this surface was
obtained by setting $k$ equal to~$2$ in the definition above of a type~2 surface of the sine kind so
that it is the same surface in all three plots. The other surface in each plot (in dashed red) is a type~1
surface of the sine kind and was considered to be the second surface in each plot. From left to right in
the three plots, the second surface was obtained by setting $k$ equal to $2$, $3$, $4$, respectively, in
the definition above of a type~1 surface of the sine kind. As already mentioned above, in the methods
presented in the previous section and in \cite{bernal5}, using gradient descent and/or dynamic programming,
the second surface is reparametrized while the first one is rotated.
\\ \smallskip\\
With $\gamma(r,t)=(r^{5/4},t)$, $(r,t) \in [0,1]\times [0,1]$,
all surfaces in the plots were discretized as described above and an elastic shape registration
of the two surfaces in each plot together with the elastic shape distance between them associated
with the registration were computed through executions of our software package, using gradient descent
and/or dynamic programming. We note that for this particular $\gamma$, the discretizations of the second
surfaces were perturbed only in the $r$ direction which made the software package more likely to succeed
just by using dynamic programming, before using gradient descent, as it is perturbations in the $r$
direction that the dynamic programming software is equipped to handle.
\\ \smallskip\\
Using dynamic programming followed by gradient descent, the results were as follows. The three elastic shape
distances computed with dynamic programming before using gradient descent, in the order of the plots from
left to right, were as follows with the first distance, as hoped for, essentially equal to zero:
0.00031, 0.3479, 0.3192. The times of execution in the same order were 28.22, 29.23, 39.97 seconds.
The computed optimal rotation matrices in the same order were
$\left( \begin{smallmatrix} -0.0008 & 1 & 0\\ 0 & 0 & 1\\ 1 & 0.0008 & 0\\ \end{smallmatrix} \right),$
$\left( \begin{smallmatrix} 0.02 & 1 & 0\\ 0 & 0 & 1\\ 1 & -0.02 & 0\\ \end{smallmatrix} \right),$
$\left( \begin{smallmatrix} 0.03 & 1 & 0\\ 0 & 0 & 1\\ 1 & -0.03 & 0\\ \end{smallmatrix} \right).$
The three elastic shape distances computed using gradient descent with the results obtained with dynamic
programming used as input, in the order of the plots from left to right, were as follows:
0.00029, 0.3421, 0.3190. The times of execution in the same order were 0.32, 0.44, 0.29 seconds.
The computed optimal rotation matrices in the same order were
$\left( \begin{smallmatrix} 0.00001 & 1 & 0\\ 0 & 0 & 1\\ 1 & -0.00001 & 0\\ \end{smallmatrix} \right),$
$\left( \begin{smallmatrix} 0.02 & 1 & 0\\ 0 & 0 & 1\\ 1 & -0.02 & 0\\ \end{smallmatrix} \right),$
$\left( \begin{smallmatrix} 0.03 & 1 & 0\\ 0 & 0 & 1\\ 1 & -0.03 & 0\\ \end{smallmatrix} \right).$
Note, perhaps because the discretizations of the second surfaces were perturbed only in the $r$
direction, dynamic programming alone appears to have produced good solutions while gradient descent
doesn't appear to have improved these solutions in a significant way.
On the other hand, using gradient descent without dynamic programming, i.e.,
gradient descent with initial solution the identity matrix and the identity diffeomorphism, the results
were as follows. The three elastic shape distances in the order of the plots from left to right were
0.25, 1.02, 0.7058. The times of execution in the same order were 3.68, 1.58, 2.36 seconds.
The computed optimal rotation matrices in the same order were
$\left( \begin{smallmatrix} 0.02 & 1 & 0\\ 0 & 0 & 1\\ 1 & -0.02 & 0\\ \end{smallmatrix} \right),$
$\left( \begin{smallmatrix} -0.02 & 1 & 0\\ 0 & 0 & 1\\ 1 & 0.02 & 0\\ \end{smallmatrix} \right),$
$\left( \begin{smallmatrix} -0.08 & 1 & 0\\ 0 & 0 & 1\\ 1 & 0.08 & 0\\ \end{smallmatrix} \right).$
Clearly these results obtained using gradient descent with initial solution the identity matrix and the
identity diffeomorphism are far from optimal.
\\ \smallskip\\
Finally, with $\gamma(r,t)=(r^{5/4},t^{5/4})$, $(r,t) \in [0,1]\times [0,1]$, again
all surfaces in the plots were discretized as described above and an elastic shape registration
of the two surfaces in each plot together with the elastic shape distance between them associated
with the registration were computed through executions of our software package, using gradient descent
and/or dynamic programming. We note that for this particular $\gamma$, the discretizations of the second
surfaces were perturbed in both the $r$ and the $t$ directions which made the dynamic programming software
less likely to succeed by itself as it is equipped to handle perturbations in the $r$ direction but not
in the $t$ direction.
\\ \smallskip\\
Using dynamic programming followed by gradient descent, the results were as follows. The three elastic shape
distances computed with dynamic programming before using gradient descent, in the order of the plots from
left to right, were as follows with the first distance close to zero but not enough:
0.0126, 0.3565, 0.3282. The times of execution in the same order were 28.76, 49.37, 40.75 seconds.
The computed optimal rotation matrices in the same order were
$\left( \begin{smallmatrix} -0.001 & 1 & 0\\ 0 & 0 & 1\\ 1 & 0.001 & 0\\ \end{smallmatrix} \right),$
$\left( \begin{smallmatrix} 0.02 & 1 & -0.00008\\ -0.0002 & 0.00009 & 1\\ 1 & -0.02 & 0.0002\\
\end{smallmatrix} \right),$
$\left( \begin{smallmatrix} 0.03 & 1 & 0.0007\\ 0.002 & -0.0008 & 1\\ 1 & -0.03 & -0.002\\
\end{smallmatrix} \right).$
The three elastic shape distances computed using gradient descent with the results obtained with dynamic
programming used as input, in the order of the plots from left to right, were as follows:
0.00302, 0.3493, 0.3234. The times of execution in the same order were 2.73, 0.77, 0.65 seconds.
The computed optimal rotation matrices in the same order were
$\left( \begin{smallmatrix} -0.0001 & 1 & 0\\ 0 & 0 & 1\\ 1 & 0.0001 & 0\\ \end{smallmatrix} \right),$
$\left( \begin{smallmatrix} 0.03 & 1 & -0.0003\\ -0.0007 & 0.0003 & 1\\ 1 & -0.03 & 0.0007\\
\end{smallmatrix} \right),$
$\left( \begin{smallmatrix} 0.03 & 1 & 0.001\\ 0.005 & -0.002 & 1\\ 1 & -0.03 & -0.005\\
\end{smallmatrix} \right).$
Note, perhaps because the discretizations of the second surfaces were perturbed in both the $r$ and $t$
directions, and the dynamic programming software is not equipped to handle perturbations in the $t$
direction, the results from dynamic programming alone, although not far from optimal were not exactly
optimal. Note as well, gradient descent did improve these results somewhat, the first distance becoming closer
to zero. On the other hand, using gradient descent without dynamic programming, i.e.,
gradient descent with initial solution the identity matrix and the identity diffeomorphism, the results
were as follows. The three elastic shape distances in the order of the plots from left to right were
0.32, 0.70, 0.6262. The times of execution in the same order were 4.62, 5.40, 3.67 seconds.
The computed optimal rotation matrices in the same order were
$\left( \begin{smallmatrix} 0.008 & 1 & -0.002\\ 0.004 & 0.002 & 1\\ 1 & -0.008 & -0.004\\ \end{smallmatrix} \right),$
$\left( \begin{smallmatrix} 0.009 & -1 & -0.002\\ 0.009 & 0.002 & -1\\ 1 & 0.009 & 0.009\\ \end{smallmatrix} \right),$
$\left( \begin{smallmatrix} 0.001 & -1 & 0.0001\\ 0.0007 & -0.0001 & -1\\ 1 & 0.001 & 0.0007\\ \end{smallmatrix} \right).$
Clearly these results obtained using gradient descent with initial solution the identity matrix and the
identity diffeomorphism are far from optimal.
\\ \smallskip\\
The next results that follow were obtained from applications of our software package on discretizations
of surfaces in $3-$dimensional space of the helicoid kind.
Given $k$, a positive integer, one type of surface to which we refer as a surface of the helicoid kind
(type~1) is defined by
\[ x(r,t)= r \cos k \pi t,\ \ \ y(r,t) = r \sin k \pi t,\ \ \ z(r,t) = k \pi t,\ \ \ (r,t) \in [0,1]\times [0,1], \]
and another one (type~2) by
\[ x(r,t) = k \pi t,\ \ \ y(r,t)= r \cos k \pi t,\ \ \ z(r,t) = r \sin k \pi t,\ \ \ (r,t) \in [0,1]\times [0,1], \]
the former a rotation of the latter by applying the rotation matrix
$\left( \begin{smallmatrix} 0 & 1 & 0\\ 0 & 0 & 1\\ 1 & 0 & 0\\ \end{smallmatrix} \right)$
on the latter, thus of similar shape.
\\ \smallskip\\
A plot depicting two surfaces (actually their boundaries) of similar shape of the helicoid
kind for $k=4$ is shown in Figure~\ref{F:helicoids}. (Note that in the plot there, the $x-$, $y-$, $z-$
axes are not always to scale relative to one another). The two surfaces being of similar shape, clearly the
elastic shape distance between them is exactly zero, and the hope was once again that the execution of our
software package applied on these two surfaces, using gradient descent and/or dynamic programming, would produce
an elastic shape distance between them equal or close to zero. The type~2 surface of the helicoid kind in the plot
(in solid blue) was considered to be the first surface in the plot. The other surface in the plot (in dashed red)
is a type~1 surface of the helicoid kind and was considered to be the second surface in the plot.
\\ \smallskip\\
With $\gamma(r,t)=(r^{5/4},t)$, $(r,t) \in [0,1]\times [0,1]$,
the two surfaces in the plot were discretized as described above and an elastic shape
registration of the two surfaces together with the elastic shape distance between them associated
with the registration were computed through the execution of our software package, using gradient
descent and/or dynamic programming. Again we note that for this particular $\gamma$, the discretization
of the second surface was perturbed only in the $r$ direction which as pointed out above made the
software package more likely to succeed just by using dynamic programming, before using gradient descent.
\begin{figure}
\centering
\begin{tabular}{cc}
\includegraphics[width=0.3\textwidth]{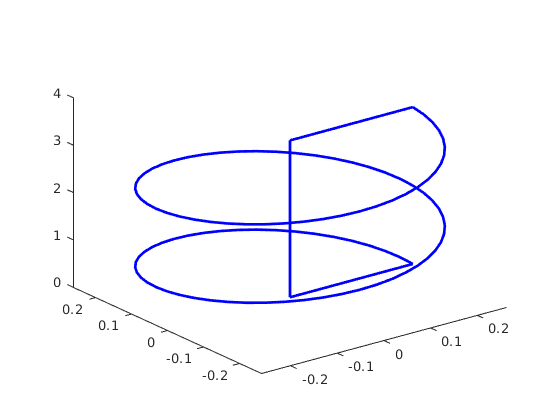}
&
\includegraphics[width=0.3\textwidth]{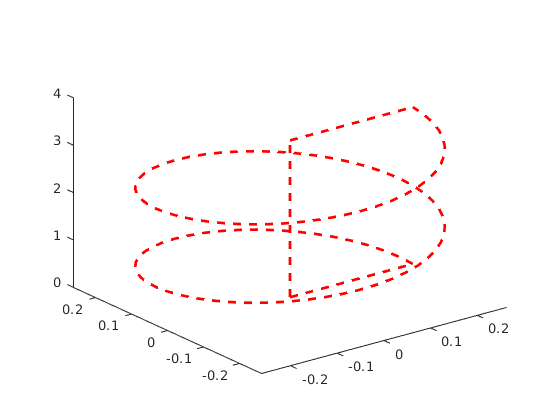}
\end{tabular}
\caption{\label{F:h123best}
For $\gamma(r,t)=(r^{5/4},t)$, $(r,t) \in [0,1]\times [0,1]$, after dynamic programming, before gradient
descent, views of boundary of rotated first surface (solid blue), and of reparametrized second surface
(dashed red).
}
\end{figure}
\\ \smallskip \\
Using dynamic programming followed by gradient descent, the results were as follows. The elastic shape
distance computed with dynamic programming before using gradient descent was 0.00019, which, as hoped for,
was close enough to zero. The time of execution was 15.32 seconds. The computed optimal rotation matrix was
$\left( \begin{smallmatrix} 0 & 1 & 0\\ 0 & 0 & 1\\ 1 & 0 & 0\\ \end{smallmatrix} \right).$
Views of the two surfaces after dynamic programming, before gradient descent,
are shown in Figure~\ref{F:h123best}. The elastic shape distance computed using gradient descent with the
results obtained with dynamic programming used as input was 0.00018. The time of execution was 0.23 seconds.
The computed optimal rotation matrix was
$\left( \begin{smallmatrix} 0 & 1 & 0\\ 0 & 0 & 1\\ 1 & 0 & 0\\ \end{smallmatrix} \right).$
Note, perhaps because the discretization of the second surface was perturbed only in the $r$
direction, dynamic programming alone appears to have produced good solutions while gradient descent
doesn't appear to have improved these solutions in a significant way.
On the other hand, using gradient descent without dynamic programming, i.e.,
gradient descent with initial solution the identity matrix and the identity diffeomorphism, the results
were as follows. The elastic shape distance was 0.11. The time of execution was 13.42 seconds.
The computed optimal rotation matrix was
$\left( \begin{smallmatrix} 0.01 & 1 & -0.06\\ -0.02 & 0.06 & 1\\ 1 & -0.01 & 0.02\\ \end{smallmatrix} \right).$
Clearly these results obtained using gradient descent with initial solution the identity matrix and the
identity diffeomorphism, although not necessarily bad, are still far from optimal.
\\ \smallskip\\
Finally, with $\gamma(r,t)=(r^{5/4},t^{5/4})$, $(r,t) \in [0,1]\times [0,1]$, again the two surfaces in the
plot were discretized as described above and an elastic shape registration of the two surfaces together with
the elastic shape distance betwen them associated with the registration were computed through the execution
of our software package, using gradient descent and/or dynamic programming. Again we note that for this
particular~$\gamma$, the discretization of the second surface was perturbed in both the $r$ and the $t$
directions which as pointed out above made the dynamic programming software less likely to succeed by itself
as it is equipped to handle perturbations in the $r$ direction but not in the $t$ direction.
\begin{figure}
\centering
\begin{tabular}{cc}
\includegraphics[width=0.3\textwidth]{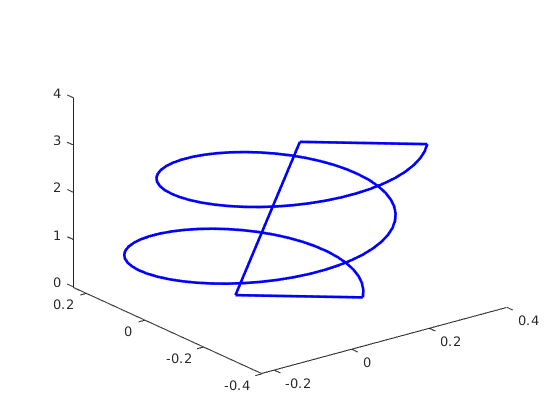}
&
\includegraphics[width=0.3\textwidth]{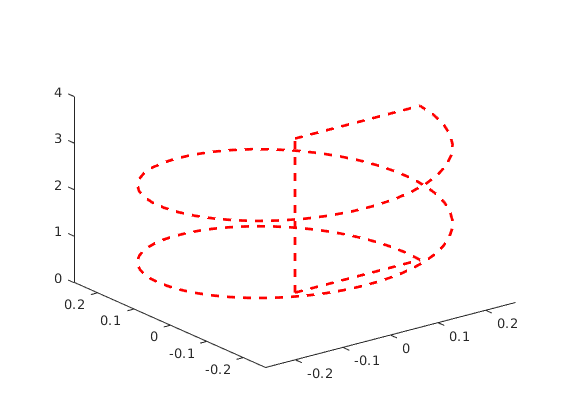}
\end{tabular}
\caption{\label{F:h312best}
For $\gamma(r,t)=(r^{5/4},t^{5/4})$, $(r,t) \in [0,1]\times [0,1]$, after dynamic programming, before gradient
descent, views of boundary of rotated first surface (solid blue), and of reparametrized second surface
(dashed red).
}
\end{figure}
\\ \smallskip \\
Using dynamic programming followed by gradient descent, the results were as follows. The elastic shape
distance computed with dynamic programming before using gradient descent was 0.0796 which
was not far from zero but not close enough. The time of execution was 19.59 seconds. The computed optimal
rotation matrix was
$\left( \begin{smallmatrix} 0.03 & 0.8 & 0.7\\ -0.03 & -0.7 & 0.8\\ 1 & -0.04 & 0.004\\
\end{smallmatrix} \right).$
Views of the two surfaces after dynamic programming, before gradient descent,
are shown in Figure~\ref{F:h312best}. The elastic shape distance computed using gradient descent with the
results obtained with dynamic programming used as input was 0.00876. The time of execution was 6.70 seconds.
The computed optimal rotation matrix was
$\left( \begin{smallmatrix} 0.01 & 1 & 0.07\\ -0.004 & -0.07 & 1\\ 1 & -0.01 & 0.003\\ \end{smallmatrix} \right).$
Views of the two surfaces after dynamic programming followed by gradient descent
are shown in Figure~\ref{F:h132best}.
Note, perhaps because the discretization of the second surface was perturbed in both the $r$
and $t$ directions, and the dynamic programming software is not equipped to handle perturbations in the $t$
direction, the results from dynamic programming alone, although not far from optimal were not exactly optimal.
Note as well, gradient descent did improve these results somewhat, the distance becoming closer to zero.
On the other hand, using gradient descent without dynamic programming, i.e.,
gradient descent with initial solution the identity matrix and the identity diffeomorphism, the results
were as follows. The elastic shape distance was 0.0943. The time of execution was 15.80 seconds.
The computed optimal rotation matrix was
$\left( \begin{smallmatrix} 0.01 & 1 & -0.05\\ -0.02 & 0.05 & 1\\ 1 & -0.01 & 0.02\\ \end{smallmatrix} \right).$
Clearly these results obtained using gradient descent with initial solution the identity matrix and the
identity diffeomorphism, although not necessarily bad, are still far from optimal.
\begin{figure}
\centering
\begin{tabular}{cc}
\includegraphics[width=0.3\textwidth]{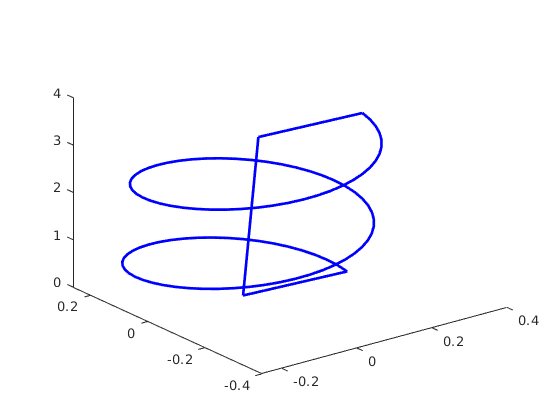}
&
\includegraphics[width=0.3\textwidth]{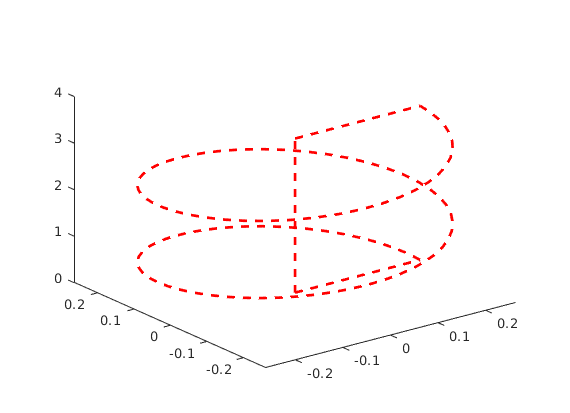}
\end{tabular}
\caption{\label{F:h132best}
For $\gamma(r,t)=(r^{5/4},t^{5/4})$, $(r,t) \in [0,1]\times [0,1]$, after dynamic programming followed by gradient
descent, views of boundary of rotated first surface (solid blue), and of reparametrized second surface
(dashed red).
}
\end{figure}
\\ \smallskip\\
The final results that follow were obtained from applications of our software package on discretizations of two
surfaces in $3-$dimensional space that are actually graphs of functions based on the product
of the cosine and sine functions.
One surface to which we refer as the type 1 surface of the cosine-sine kind is defined by
\[ x(r,t)=r,\ \ \ y(r,t)=t,\ \ \ z(r,t)=(\cos0.5\pi r)(\sin0.5\pi t),\ \ \ (r,t) \in [0,1]\times [0,1], \]
and the other surface to which we refer as the type~2 surface of the cosine-sine kind is defined by
\[ x(r,t)=(\cos0.5\pi r)(\sin0.5\pi t),\ \ \ y(r,t)=r,\ \ \ z(r,t)=t,\ \ \ (r,t) \in [0,1]\times [0,1], \]
the former a rotation of the latter by applying the rotation matrix
$\left( \begin{smallmatrix} 0 & 1 & 0\\ 0 & 0 & 1\\ 1 & 0 & 0\\ \end{smallmatrix} \right)$
on the latter, thus of similar shape.
\\ \smallskip\\
A plot depicting the two surfaces (actually their boundaries) of the cosine-sine
kind is shown in Figure~\ref{F:cos_sin}. (Note that in the plot there, the $x-$, $y-$, $z-$
axes are not always to scale relative to one another). The two surfaces being of similar shape, clearly the
elastic shape distance between them is exactly zero, and the hope was once again that the execution of our
software package applied on these two surfaces would produce an elastic shape distance between them equal
or close to zero. The type~2 surface of the cosine-sine kind in the plot (in solid blue) was considered to be
the first surface in the plot. The other surface in the plot (in dashed red)
is the type~1 surface of the cosine-sine kind and was considered to be the second surface in the plot.
\begin{figure}
\centering
\begin{tabular}{ccc}
\includegraphics[width=0.43\textwidth]{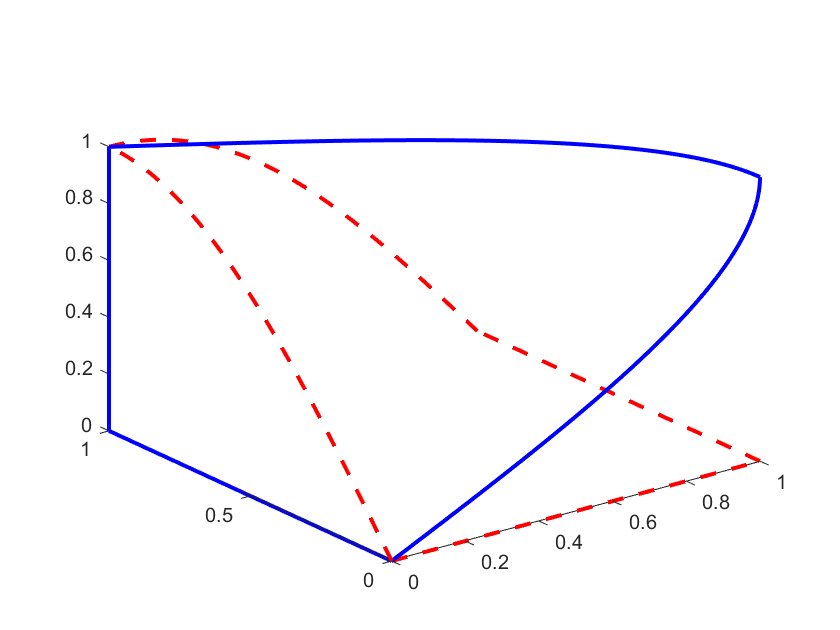}
\end{tabular}
\caption{\label{F:cos_sin}
Boundaries of the two surfaces of the cosine-sine kind, the type~$1$ surface in dashed red,
the type~$2$ surface in solid blue.
}
\end{figure}
\\ \smallskip\\
With $\gamma(r,t)=(r^{5/4},t)$, $(r,t) \in [0,1]\times [0,1]$,
the two surfaces in the plot were then discretized as described above and an elastic shape
registration of the two surfaces and the elastic shape distance between them associated with the
registration were then computed through the execution of our software package, using gradient
descent and/or dynamic programming. Again we note that for this particular $\gamma$, the discretization
of the second surface was perturbed only in the $r$ direction which as pointed out above made the
software package more likely to succeed just by using dynamic programming before using gradient descent.
\begin{figure}
\centering
\begin{tabular}{cc}
\includegraphics[width=0.46\textwidth]{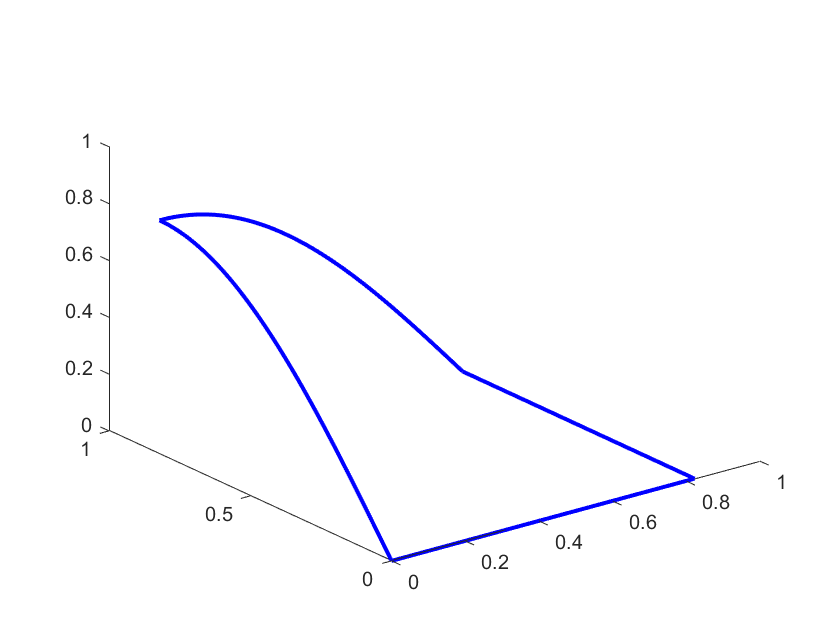}
&
\includegraphics[width=0.46\textwidth]{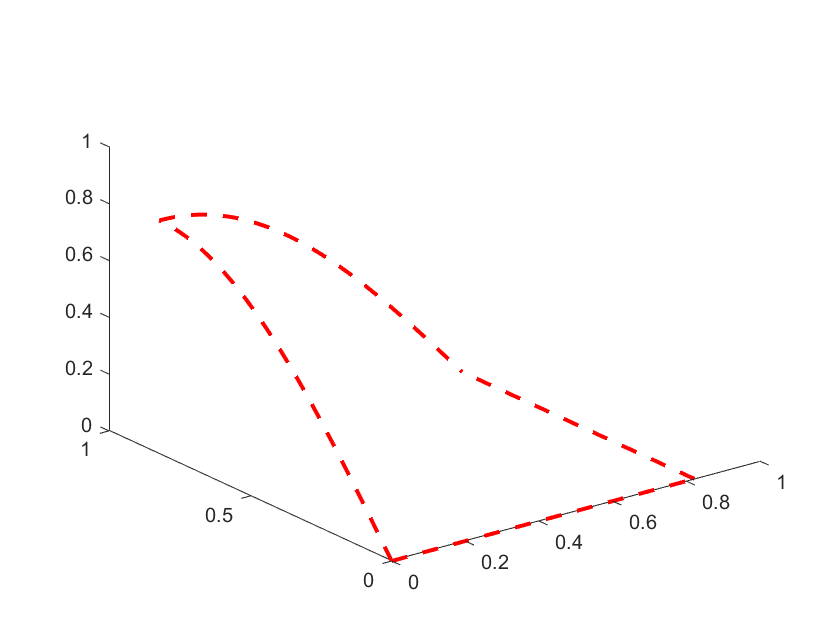}
\end{tabular}
\caption{\label{F:c123best}
For $\gamma(r,t)=(r^{5/4},t)$, $(r,t) \in [0,1]\times [0,1]$, after dynamic programming, before gradient
descent, views of boundary of rotated first surface (solid blue), and of reparametrized second surface
(dashed red).
}
\end{figure}
\\ \smallskip\\
Using dynamic programming followed by gradient descent, the results were as follows. The elastic shape
distance computed with dynamic programming before using gradient descent was 0.00021, which, as hoped for,
was close enough to zero. The time of execution was 22.29 seconds. The computed optimal rotation matrix was
$\left( \begin{smallmatrix} -0.001 & 1 & 0.0009\\ -0.001 & -0.0009 & 1\\ 1 & 0.001 & 0.001\\
\end{smallmatrix} \right).$
Views of the two surfaces after dynamic programming, before gradient descent, are shown in
Figure~\ref{F:c123best}. The elastic shape distance computed using gradient descent with the
results obtained with dynamic programming used as input was 0.00019. The time of execution was 0.23 seconds.
The computed optimal rotation matrix was
$\left( \begin{smallmatrix} -0.0006 & 1 & 0.0004\\ -0.0005 & -0.0004 & 1\\ 1 & 0.0006 & 0.0005\\
\end{smallmatrix} \right).$
Note, perhaps because the discretization of the second surface was perturbed only in the $r$
direction, dynamic programming alone appears to have produced good solutions while gradient descent
doesn't appear to have improved these solutions in a significant way.
On the other hand, using gradient descent without dynamic programming, i.e.,
gradient descent with initial solution the identity matrix and the identity diffeomorphism, the results
were as follows. The elastic shape distance was 0.47. The time of execution was 3.12 seconds.
The computed optimal rotation matrix was
$\left( \begin{smallmatrix} 0.03 & 1 & -0.1\\ -0.07 & 0.1 & 1\\ 1 & -0.02 & 0.07\\ \end{smallmatrix} \right).$
Clearly these results obtained using gradient descent with initial solution the identity matrix and the
identity diffeomorphism are far from optimal.
\\ \smallskip\\
Finally, with $\gamma(r,t)=(r^{5/4},t^{5/4})$, $(r,t) \in [0,1]\times [0,1]$, again the two surfaces in the
plot were discretized as described above and an elastic shape registration of the two surfaces together with
the elastic shape distance betwen them associated with the registration were computed through the execution
of our software package, using gradient descent and/or dynamic programming. Again we note that for this
particular~$\gamma$, the discretization of the second surface was perturbed in both the $r$ and the $t$
directions which as pointed out above made the dynamic programming software less likely to succeed by itself
as it is equipped to handle perturbations in the $r$ direction but not in the $t$ direction.
\begin{figure}
\centering
\begin{tabular}{cc}
\includegraphics[width=0.46\textwidth]{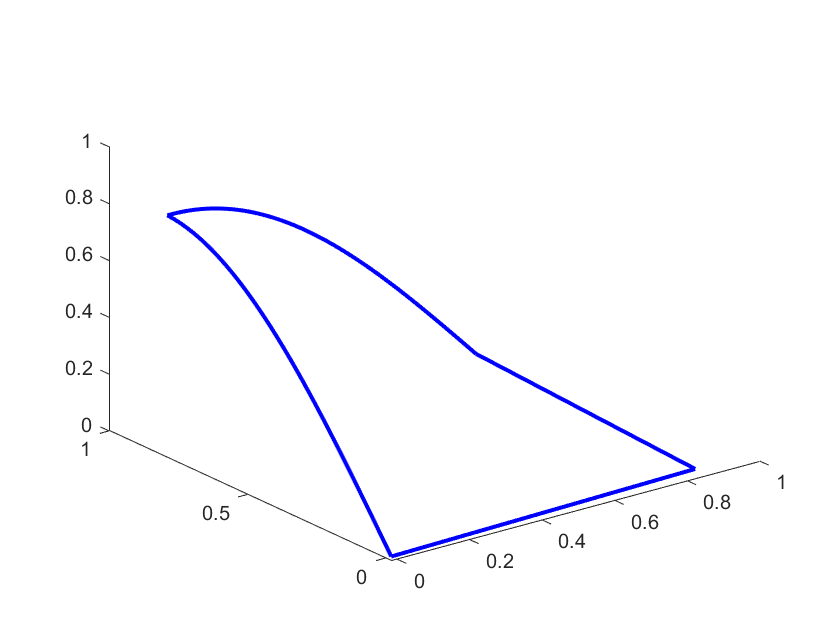}
&
\includegraphics[width=0.46\textwidth]{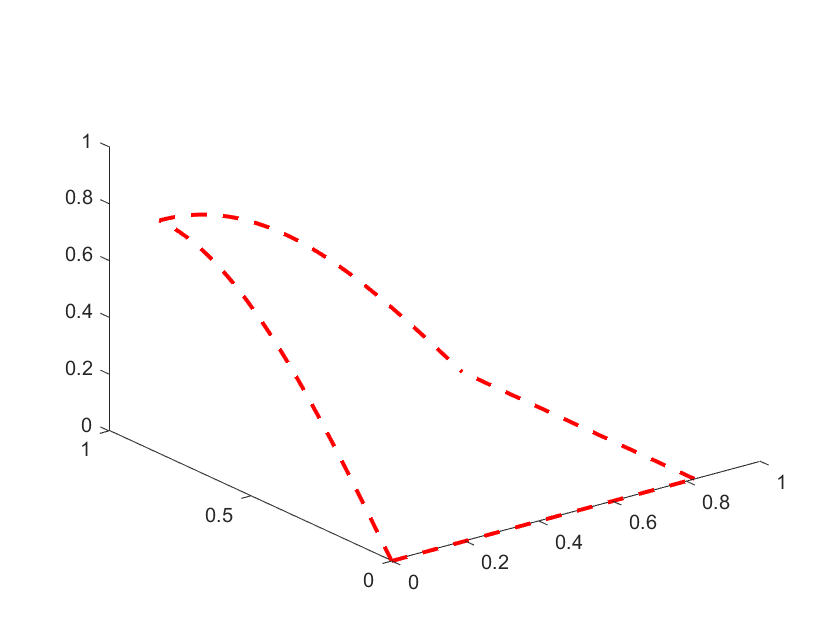}
\end{tabular}
\caption{\label{F:c312best}
For $\gamma(r,t)=(r^{5/4},t^{5/4})$, $(r,t) \in [0,1]\times [0,1]$, after dynamic programming, before gradient
descent, views of boundary of rotated first surface (solid blue), and of reparametrized second surface
(dashed red).
}
\end{figure}
\\ \smallskip \\
Using dynamic programming followed by gradient descent, the results were as follows. The elastic shape
distance computed with dynamic programming before using gradient descent was 0.0143 which
was not far from zero but not close enough. The time of execution was 23.54 seconds. The computed optimal
rotation matrix was
$\left( \begin{smallmatrix} -0.04 & 1 & 0.03\\ -0.04 & -0.03 & 1\\ 1 & 0.04 & 0.04\\
\end{smallmatrix} \right).$
Views of the two surfaces after dynamic programming, before gradient descent,
are shown in Figure~\ref{F:c312best}. The elastic shape distance computed using gradient descent with the
results obtained with dynamic programming used as input was 0.0033. The time of execution was 2.11 seconds.
The computed optimal rotation matrix was
$\left( \begin{smallmatrix} -0.02 & 1 & 0.008\\ -0.02 & -0.008 & 1\\ 1 & 0.02 & 0.02\\ \end{smallmatrix} \right).$
Views of the two surfaces after dynamic programming followed by gradient descent are shown in
Figure~\ref{F:c132best}.
Note, perhaps because the discretization of the second surface was perturbed in both the $r$
and $t$ directions, and the dynamic programming software is not equipped to handle perturbations in the $t$
direction, the results from dynamic programming alone, although not far from optimal were not exactly optimal.
Note as well, gradient descent did improve these results somewhat, the distance becoming closer to zero.
On the other hand, using gradient descent without dynamic programming, i.e.,
gradient descent with initial solution the identity matrix and the identity diffeomorphism, the results
were as follows. The elastic shape distance was 0.4791. The time of execution was 3.28 seconds.
The computed optimal rotation matrix was
$\left( \begin{smallmatrix} 0.01 & 1 & -0.1\\ -0.08 & 0.1 & 1\\ 1 & -0.003 & 0.08\\ \end{smallmatrix} \right).$
Clearly these results obtained using gradient descent with initial solution the identity matrix and the
identity diffeomorphism are far from optimal.
\begin{figure}
\centering
\begin{tabular}{cc}
\includegraphics[width=0.46\textwidth]{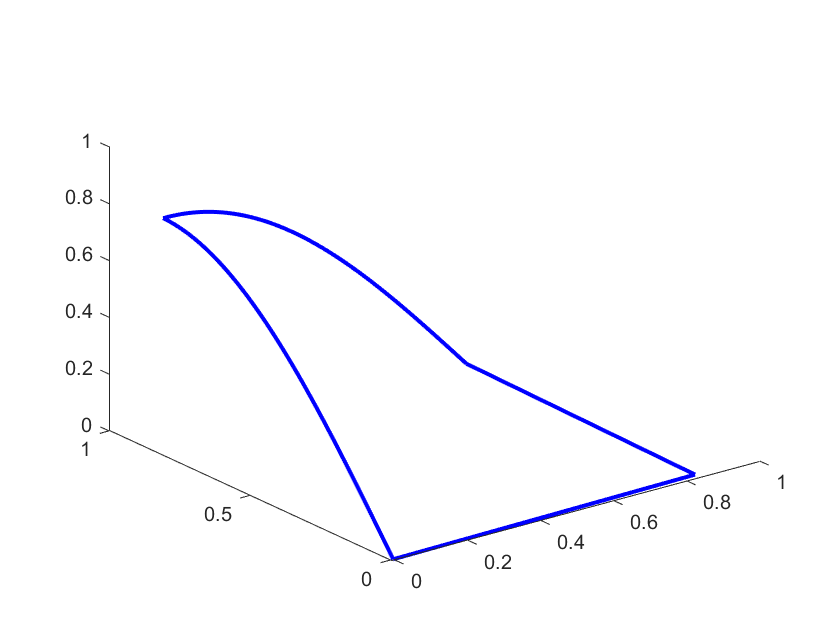}
&
\includegraphics[width=0.46\textwidth]{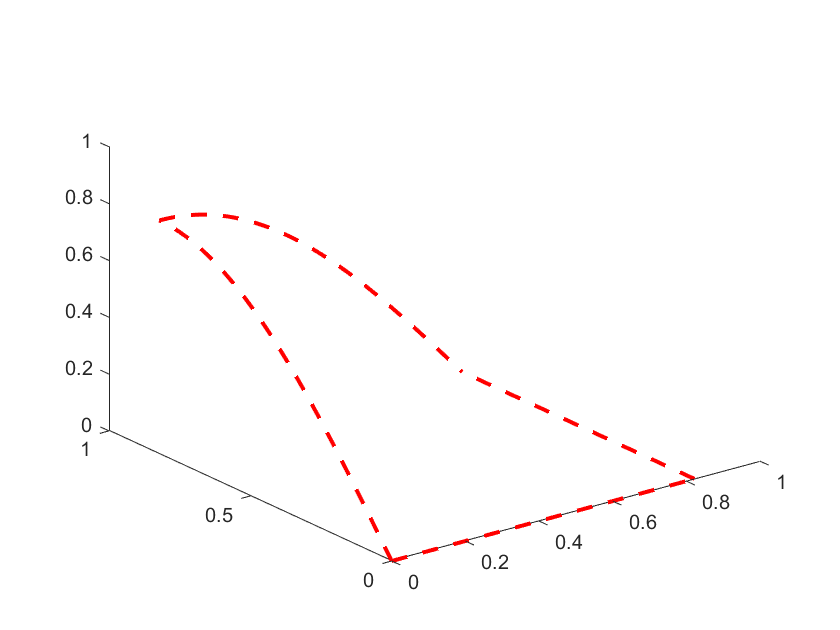}
\end{tabular}
\caption{\label{F:c132best}
For $\gamma(r,t)=(r^{5/4},t^{5/4})$, $(r,t) \in [0,1]\times [0,1]$, after dynamic programming followed by gradient
descent, views of boundary of rotated first surface (solid blue), and of reparametrized second surface
(dashed red).
}
\end{figure}
\\ \smallskip\\ \noindent
{\bf\large Summary}
\\ \smallskip\\
In this paper we have presented results from computing the elastic shape registration of two simple
surfaces in $3-$dimensional space and the elastic shape distance between them with an algorithm based
on a gradient descent approach for reparametrizing one of the surfaces, using as the input initial
solution to the algorithm the rotation and reparametrization computed with the algorithm based on
dynamic programming presented in~\cite{bernal5} for reparametrizing one of the surfaces to obtain
a partial elastic shape registration of the surfaces. The gradient descent approach used to obtain
our results is a generalization to surfaces in $3-$dimensional space of the gradient descent approach
for reparametrizing one of two curves in the plane when computing the elastic shape distance between
them as presented in \cite{srivastava}. We have described and justified the approach for curves as it
is done in \cite{srivastava}, and have presented and justified its generalization to surfaces in
$3-$dimensional space. Our algorithm based on gradient descent and dynamic programming as just described
has been implemented in the form of a software package written in Matlab with the exception of the
dynamic programming routine which is written in Fortran but executed as a Matlab mex file. The results
we have presented in this paper were obtained from applications of the software package on
discretizations of the three kinds of surfaces in $3-$dimensional space identified in \cite{bernal5}
as surfaces of the sine, helicoid and cosine-sine kind. Some of these results verify that dynamic
programming alone as implemented produces good results essentially optimal for surfaces whose
parametrizations have been perturbed only in the $x-$direction of the plane as it is perturbations in
this direction that dynamic programming as implemented is equipped to handle. On the other hand, other
results verify that dynamic programming alone as implemented produces results not far from
optimal but still not optimal for surfaces whose parametrizations have been perturbed in the
$y-$direction of the plane as well. For this situation, other results show that gradient descent
using as input the results obtained with dynamic programming produces results closer to
optimal than dynamic programming alone as implemented. However some results show that gradient
descent alone without dynamic programming, i.e., gradient descent with initial solution the identity
matrix and the identity diffeomorphism produces results far from optimal almost every time.

\end{document}